# Multistep reversible excitation transfer in a multicomponent rigid solution: II. Modeling the dynamics of radiationless transfer as a time-resolved Markov chain

**Józef Kuśba**[1]
*Independent researcher. Retired at Faculty of Applied Physics and Mathematics, Gdańsk University of Technology, Gdańsk, Poland*

**Abstract**
To determine the effect of nonradiative excitation energy transfer on the fluorescence of a rigid multicomponent solution, a new analytical method was developed by treating this transfer as a time-resolved Markov chain (TRMC). In the TRMC approach, we assume that the Markov chain under consideration is governed by bivariate joint probability mass-density functions. One of the random variables is discrete and represents the state number to which the process passes at a given step, while the other random variable is continuous and determines the moment in time at which this transition occurs. In general, the time distributions of this second variable can be arbitrary, continuous or discrete, and not just exponential, as required by the method known as continuous time Markov chains (CTMC). The agreement between the basic expressions of the TRMC method and the analogous expressions of the renewal theory has been demonstrated. The correctness of the TRMC method is confirmed by the fact that the time courses of fluorescence intensities calculated by this method agree with those calculated using ordinary analytical methods. In the section on calculating the quantum yields of individual components, the suitability of a method known as discrete time Markov chains (DTMC) was found. However, we argue that the DTMC method does not refer to time and propose to rename it as time-unspecified Markov chain (TUMC). The results generated by TRMC, when integrated over time, become equivalent to those generated by TUMC. The fluorescence cases of binary and ternary solutions are discussed in detail.

## Abbreviations

| | | | |
|---|---|---|---|
| CTMC | continuous time Markov chain | RNET | reversible nonradiative energy transfer |
| DTMC | discrete time Markov chain | RT | renewal theory |
| FRET | Förster resonance energy transfer | SCDF | subnormalized cumulative distribution function |
| MCS | multicomponent solution | SPDF | subnormalized probability density function |
| MEE | molecular electronic excitation | TRMC | time-resolved Markov chain |
| NET | nonradiative excitation transfer | TUMC | time-unspecified Markov chain |

## Symbols and notation

### Latin letters

| | |
|---|---|
| $a_{ij}$ | distance of the closest approach of the $i$th and $j$th component molecules. Eqs. (13), (36) |
| $c_i$ | concentration of the $i$th component molecules. Eq. (13) |
| $E(t)$ | SPDF vector of effective fluorescence. Eq. (146) |
| $\mathcal{F}_{\mathrm{M}i}^{(1)}(t)$ | cumulative distribution function of one-step events on $i$th component molecules. Eq. (18) |
| $f_{\mathrm{F}}(x)$ | special function. Eq. (46) |
| $g(t;\gamma,\tau)$ | auxiliary function. Eq. (41) |
| $g_i(t)$ | auxiliary function. Eq. (184) |
| $h$ | time segment, Eq. (49) |
| $h_i(t)$ | auxiliary function. Eq. (185) |
| $H(t)$ | Heaviside step function. Eq. (118) |
| $I_\nu$ | identity matrix of dimension $\nu\times\nu$. Eq. (69) |
| $J$ | state number random variable. Sec. 3.3 |
| $k_{ij}(r)$ | bimolecular transfer rate. Eqs. (13), (35) |

[1] ORCID: 0000-0002-9697-0751. Electronic mail: jozkusba@pg.edu.pl

$\mathbb{K}_{ij}(t)$ distance-averaged transfer rate. Eqs. (13), (36), (38)

$k_{\mathrm{r}}$ matrix of radiative rate constants. Eq. (11)

$k_{\mathrm{nr}}$ matrix of nonradiative rate constants. Eq. (12)

$\mathcal{L}$ Laplace transform operator. Eq. (71)

$m$ the number of terminating states taken into account. Sec. 3.2

$n$ the number of components in the MCS under consideration. Sec. 1.2

$O_{m\times n}$ zero matrix of dimension $m\times n$ . Eq. (63)

$P$ transition probability matrix. Eq. (98)

$P^{(\mu)}$ $\mu$-step transition probability matrix. Eqs. (63), (65)

$\mathbb{P}(t)$ matrix of intensities of MEE visits to individual states. Eq. (69)

$\mathbb{P}^{(\mu)}(t)$ $\mu$-step transition SPDF matrix. Eqs. (50), (59), (66), (68)

$\mathcal{P}(t)$ residence probability matrix. Eqs. (113), (116)

$\mathcal{P}^{(1)}(t)$ one-step residence probability matrix. Eq. (114)

$\mathcal{P}_{\mathrm{M}}(t)$ residence probability matrix of the MEE. Eq. (120)

$\mathcal{P}_{\Psi}(t)$ residence probability matrix of the converted MEE. Eq. (120)

$\mathcal{P}_{\mathrm{M}}^{(1)}(t)$ matrix of probabilities of one-step MEE residence on individual components. Eqs. (15), (118)

$Q(t)$ renewal kernel. Eq. (127)

$R(t)$ renewal matrix. Eq. (129)

$R_i^{(1)}(t)$ marginal one-step distribution of the random variable $T$ in the state $i$. Eq. (53)

$R_{0ij}$ Förster radius for NET from $i$th component molecules to $j$th component molecules. Eq. (35)

$s$ Laplace variable. Eq. (71)

$\mathcal{S}$ state space. Sec. 3.3

$t$ time. Eq. (1)

$t_{\mu}$ moment of the beginning of step numbered $\mu$. Eq. (51)

$T$ time random variable. Sec. 3.3

$\mathcal{T}$ time space. Sec. 3.3

$W(t)$ auxiliary vector. Eq. (151)

$x_i(s)$ NET parameter. Eq. (77)

Greek letters

$\delta(t)$ Dirac delta function. Eq. (69)

$\delta_{ij}$ Kronecker delta function. Eq. (50)

$\gamma_{ij}$ reduced concentration for NET from $i$th to $j$th component. Eq. (37)

$\Gamma_i$ sum of reduced concentrations for NET from the $i$th component. Eq. (40)

$\zeta$ probability matrix of internal conversion, Eq. (103)

$\zeta^{(1)}$ probability matrix of one-step internal conversion. Eq. (33)

$Z^{(1)}(t)$ SPDF matrix of one-step internal conversion. Eq.(27), (44)

$\theta$ matrix of numbers of MEE transfers. Eq. (100)

$\theta^{(1)}$ probability matrix of one-step NET. Eqs. (31), (63)

$\Theta(t)$ matrix of intensities of MEE transfer. Eq. (93)

$\Theta^{(1)}(t)$ SPDF matrix of one-step NET. Eqs. (23), (59)

$\mu$ index of a stochastic process, step number. Eq. (4)

$\mathcal{M}$ set of values of $\mu$. Eq. (4)

$\nu$ total number of states of the chain. Sec. 3.2

$\rho_{\mathrm{M}i}^{(1)}(t)$ PDF of one-step deexcitation of the $i$th component molecules. Eq. (19)

$\tau_0$ matrix of fluorescence lifetimes at very low concentrations. Eq. (7)

$\phi$ probability matrix of photon emission. Eq. (2)

$\phi_0$ matrix of absolute quantum yields. Eq. (8)

$\phi^{(1)}$ probability matrix of one-step photon emission. Eq. (32)

| | |
|---|---|
| $\Phi(t)$ | SPDF matrix of photon emission. Eq. (1) |
| $\Phi^{(1)}(t)$ | SPDF matrix of one-step photon emission. Eq. (25) |
| $\psi$ | probability matrix of deexcitation. Eqs. (100), (102) |
| $\psi^{(1)}$ | probability matrix of one-step deexcitation. Eq. (63), (64) |
| $\Psi(t)$ | SPDF matrix of deexcitation. Eq. (93), (94) |
| $\Psi^{(1)}(t)$ | SPDF matrix of one-step deexcitation. Eq. (59), (60) |

# 1 Introduction

The phenomenon of non-radiative excitation transfer (NET) between fluorescent particles being at sufficiently small distances from each other has been studied for more than 100 years. The earliest works related to this issue concerned collisional transfer in gases [1-4]. This type of transfer occurs in systems where momentary direct contact of molecules with each other is possible. Later, as a result of the work of Förster [5,6] and Dexter [7], fluorescence studies of molecular systems in which the NET processes take place at a distance were strongly developed. Direct contact between the molecules involved in the energy transfer is not necessary in such systems. This type of NET allows to describe the properties of fluorescence emitted by rigid systems, but to some extent also by liquid systems. The condition for the appearance of the mechanism of energy transfer at a distance is a greater or lesser overlap of the emission spectrum of the excitation energy donor and absorption spectrum of the excitation energy acceptor. Such transfer can occur not only between different fluorescent molecules, but also between different fluorescent chromophore groups located on the same macromolecule [8,9].

In this work, in order to shorten our considerations, we will limit ourselves to describing the fluorescence properties of rigid solutions of various types of single-chromophore molecules. For the same reasons, in our considerations we will omit single-component solutions as relatively simple, the properties of which are easily described from the expressions obtained for binary solutions. We will assume that the solvent used is non-fluorescent, and that each of the solute types of fluorescent molecules is a separate component of the solution.

## 1.1 The state of NET description in multicomponent solutions to date

Most work on the theoretical description of the fluorescence of systems containing more than one component was devoted to two-component systems. The initial work dealt with the situation when only one component called the excitation energy donor was directly excited, and then the excitation energy could be radiationlessly transferred to the second component called the excitation energy acceptor. For such conditions, the quantum yield and fluorescence decay function of the donor were calculated, assuming a one-step transfer of molecular excitation energy (MEE) to the acceptor. For dipole-dipole interaction, the work of Förster [6] is of primary importance here, and for exchange interaction the work of Inokuti and Hirayama [10]. The effect of MEE migration in the donor system on these quantities was considered in [11] and [12]. The effect of back transfer, that is, the additional transfer of MEE from the acceptor to the donor, on fluorescence yield was considered in [13]. In non-viscous solutions, the presence of material diffusion further increases the intensity of radiationless MEE transfer between molecules. The effect of diffusion on the decay of donor fluorescence was considered in the works [14] and [15]. The effect of diffusion and migration in the donor system on the fluorescence yield of the donor was considered in the work [16], and the effect of diffusion and back transfer on the decay of the fluorescence intensity of the donor and acceptor was discussed in [17]. Consideration of the effect of correlation between the successive NETs was undertaken in the works [18-21]. It should be noted that if back transfer is included in the description of the fluorescence of a binary system, such terms as excitation energy donor and/or excitation energy acceptor become unclear and thus of little use. In such a case, it is better to number the individual components of the system and relate the theoretical results obtained to the components with specific numbers. This problem becomes especially important for systems where the number of components is greater than two.

In the first works on the description of the fluorescence intensity of ternary solutions, it was assumed that the fluorescence efficiencies of individual components as well as the MEE transfer efficiencies between components were parameters to be determined by matching the postulated theoretical expressions with experimental data [22,23]. To reduce the number of parameters determined in this way, consideration was limited to cases where NET between components is unidirectional and no back transfer is possible. In the case of multicomponent systems, however, such a way of determining the values of emission and transfer parameters leads to an ill-conditioned system of equations. The papers [24-26] show how the values of these parameters can be determined from independent measurements and using expressions obtained from extending to ternary systems the method developed in [11]. Considerations of MEE transfer in binary and ternary systems were also given in [27-29] when discussing the use of these systems as active substances in dye lasers. An algorithm for decomposing the fluorescence spectrum of a ternary mixture assuming linear Stern-Volmer extinction kinetics is given in [30].

So far, attempts to describe the effect of NET on the fluorescence properties of multi-component solutions (MCS), that is, solutions containing any number of different fluorescent components, have been made in two papers. In [31], expressions describing the quantum yields of the individual components of a multicomponent solution adequate for the situation when in each pair of solution components, NET takes place in only one direction and has a hopping character. Expressions describing, under analogous conditions, both the quantum yields and fluorescence decay of the individual components taking into account MEE back-transfer were obtained in [32].

### 1.2 Aim of the study. Basic parameters describing the effect of NET on MCS fluorescence

In this work, as in [33], the object of our consideration is the fluorescence properties of MCS containing $n$ different fluorophores dissolved in an optically inactive solvent. We assume that the individual fluorophores do not react with each other, and that they all exhibit single-exponential fluorescence decays at low concentrations. Furthermore, we assume that in the described system the effect of material diffusion on NET efficiency can be neglected, and that individual fluorophores of MCS do not emit delayed fluorescence and/or phosphorescence. From the point of view of the properties of the fluorescence emitted by MCS, the processes of excitation energy transfer between the individual components are of fundamental importance. Radiative transfer processes were discussed in detail in [33], while the purpose of this work is to discuss the effect of radiationless transfer on the temporal distribution of fluorescence emitted by individual solution components. In [33], it was shown that in the latter aspect the quadratic, $n \times n$ matrix $\Phi(t)$ is important

$$\Phi(t) = \left[ \Phi_{ij}(t) \right]_{n \times n} \tag{1}$$

The individual elements $\Phi_{ij}(t)$ of this matrix describe the probability densities of the emission of a light quantum by the $j$th component molecule at time $t$ as a result of the fact that at $t = 0$ the excitation light generated the excited state of the $i$th component molecule. The $\Phi(t)$ matrix is therefore of primary importance for describing the time dependence of the intensity of the observed fluorescence after pulse excitation. Note that the row number of the $\Phi(t)$ matrix specifies the number of the component to which MEE is delivered at $t = 0$, and the column number specifies the number of the component whose fluorescence is observed.

To describe the intensity of fluorescence observed after excitation with light of constant intensity, it is sufficient to use the time-independent $n \times n$ element $\phi$ matrix

$$\phi = \left[ \phi_{ij} \right]_{n \times n} \tag{2}$$

whose elements $\phi_{ij}$ are the probabilities of emitting a quantum of light by the $j$th component molecule if the excitation light generated the excited state of the $i$th component molecule. From the definitions of $\phi_{ij}$ and $\Phi_{ij}(t)$, it follows that the $\phi$ matrix can be obtained by integrating the $\Phi(t)$ matrix over time from zero to infinity

$$\phi = \int_0^\infty \Phi(t)\, dt \tag{3}$$

It is worth noting that the $\phi$ and $\Phi(t)$ matrices refer to the probabilities of only those MEE deactivation processes that lead to light quantum emission, and do not refer, for example, to the probabilities of radiationless deactivation processes. Therefore, in general, the elements of the $\phi$ matrix do not satisfy the normalization condition, and usually there is $\sum_{j=1}^{n} \phi_{ij} < 1$. The main part of the paper is devoted to a discussing of how to find the elements of the $\Phi(t)$ and/or $\phi$ matrix that characterize the fluorescence of rigid MCS.

### 1.3 NET in MCS as a multi-step process

In determining the values of the elements of the $\phi$ and $\Phi(t)$ matrices, we will assume that the time evolution of a MEE in MCS is multistep. Each of these steps begins when an MEE localizes to a given fluorescent molecule and continues until that MEE is transferred to another fluorescent molecule, or is converted to other types of energy. In our discussion, we will numerate the successive steps of a given MEE using the index $\mu$, which we will call step number. We will assume that the set $\mathcal{M}$ of values of $\mu$ is countable and discrete

$$\mathcal{M} = \{\mu \in \mathbb{N}\} = \{1, 2, \ldots\} \tag{4}$$

We assign $\mu = 1$ to the step that begins with the excitation of the molecule directly with an external light beam, we assign $\mu = 2$ to the step in which MEE was delivered to the molecule as a result of one transfer from the directly excited molecule, we assign $\mu = 3$ to the step in which MEE was delivered to the molecule as a result of two transfers from the directly excited molecule, and so on. We will also assume that the processes for the transfer and conversion of MEE are homogeneous, that is, their probabilities are independent of the current value of the $\mu$ step number. Under such conditions, the multistep MEE transfer kinetics is a simple composite of repeated identical one-step kinetics.

## 2 One-step NET kinetics in MCS

### 2.1 Rates of elementary processes occurring in MCS after optical excitation

Absorption of photons from the excitation beam causes the appearance of excited molecules in the MCS under study. Our considerations will apply to situations in which the number of excited molecules is small compared to the total number of fluorescent molecules in the illuminated part of the solution. This condition is usually met when using conventional excitation light sources. If a laser is used, its intensity should be weakened accordingly. The excitation energy remains on a given molecule for a shorter or longer period of time, but never permanently. This is because for each excited molecule there are a certain number of de-excitation pathways [34]. For simplicity, we will consider only three of them: fluorescence emission, internal conversion, and energy transfer to other molecules in the system.

#### 2.1.1 Case of isolated fluorescent molecules

The simplest description of MCS fluorescence applies when the concentrations of the components are so low that MEE transfer between any types of fluorescent molecules can be neglected. Then the off-diagonal elements of the $\Phi(t)$ matrix become zeros, so that the matrix reduces to the diagonal matrix $\Phi_0(t)$

$$\Phi_0(t) = \operatorname{diag}\left( \Phi_{01}(t), \ldots, \Phi_{0n}(t) \right) \tag{5}$$

In our considerations, we will assume that the functions $\Phi_{0i}(t)$ decrease exponentially with time and have the form

$$\Phi_{0i}(t) = k_{\mathrm{r}i} \exp\left(-t/\tau_{0i}\right) \tag{6}$$

where $k_{\mathrm{r}i}$ is the rate constant for fluorescence emission, and $\tau_{0i}$ is the average fluorescence lifetime of the *i*th component molecules. Analogous to (5), the $\tau_{0i}$ times can be treated as elements of the $\tau_0$ matrix

$$\tau_0 = \mathrm{diag}\left(\tau_{01},\ldots,\tau_{0n}\right) \tag{7}$$

Similarly, at low component concentrations, the quantum yield matrix $\phi$ given by Eq. (2) can be replaced by a diagonal matrix $\phi_0$ of the form

$$\phi_0 = \mathrm{diag}\left(\phi_{01},\ldots,\phi_{0n}\right) \tag{8}$$

where $\phi_{0i}$ is the absolute fluorescence quantum yield of the *i*th component. If we assume that for the molecules of the *i*th component, the rate constant for internal conversion is $k_{\mathrm{nr}i}$, then the magnitudes $\tau_{0i}$ and $\phi_{0i}$ depend on the rate constants $k_{\mathrm{r}i}$ and $k_{\mathrm{nr}i}$ according to the equations

$$\tau_{0i} = \frac{1}{k_{\mathrm{r}i} + k_{\mathrm{nr}i}} \tag{9}$$

$$\phi_{0i} = \frac{k_{\mathrm{r}i}}{k_{\mathrm{r}i} + k_{\mathrm{nr}i}} \tag{10}$$

The quantities $\tau_{0i}$ and $\phi_{0i}$ can be determined experimentally by measuring the fluorescence of each MCS component separately. Then, the diagonal matrices $k_{\mathrm{r}} = \mathrm{diag}\left(k_{\mathrm{r}1},\ldots,k_{\mathrm{r}n}\right)$ and $k_{\mathrm{nr}} = \mathrm{diag}\left(k_{\mathrm{nr}1},\ldots,k_{\mathrm{nr}n}\right)$ containing the rate constants $k_{\mathrm{r}i}$ and $k_{\mathrm{nr}i}$, respectively, can be calculated using the expressions

$$k_{\mathrm{r}} = \phi_0\, \tau_0^{-1} \tag{11}$$

$$k_{\mathrm{nr}} = \left(I_n - \phi_0\right)\tau_0^{-1} = \left(I_n - \phi_0\right)\phi_0^{-1}\, k_{\mathrm{r}} \tag{12}$$

where $I_n$ is the identity matrix of dimension $n \times n$.

### *2.1.2 Case of mutually interacting fluorescent molecules*

If the distance between the fluorescent molecules of the system under consideration is not very large, then their mutual interaction becomes important. Of primary importance here is the dipole-dipole interaction and the exchange interaction. As a result of this interaction, NET from one molecule to another becomes possible [34]. For now, we will not go into detail about the nature of the interaction of fluorescent molecules, and we will only assume that in our system, the interaction of molecules can lead to NET defined by rate constants $k_{ij}(r)$, which depend on the type of interacting molecules and on the distance *r* between them. Rate $k_{ij}(r)$ describes homotransfer when $j = i$, and heterotransfer when $j \neq i$.

Finding expressions to describe macroscopically observed fluorescence intensity decays and/or observed fluorescence yields, by directly using rates $k_{ij}(r)$ is difficult. Therefore, many authors use rates that result from averaging bimolecular rates $k_{ij}(r)$ over all possible distances between interacting molecules. Such averaged rates (sometimes called second order rates) can be calculated from the expression [6,35,36]

$$\mathbb{k}_{ij}(t) = 4\pi\, c_j \int_{a_{ij}}^{\infty} r^2\, k_{ij}(r) \exp\left[-k_{ij}(r)\, t\right] dr \tag{13}$$

where $c_j$ is the concentration of the component *j*, and $a_{ij}$ is the distance of the closest approach of the *i*th and *j*th component molecules. In considering the effect of NET on MCS fluorescence, it is also convenient to use the $\mathbb{k}(t)$ matrix formed from rates $\mathbb{k}_{ij}(t)$

$$\mathbb{k}(t) = \left[\mathbb{k}_{ij}(t)\right]_{n\times n} \tag{14}$$

The technique of using such time-dependent rate coefficients is widely discussed in the paper [37]. The second order rates defined by Eq. (13) describe well the energy transfer in systems where no migration of excitation energy occurs between molecules of the same component. However, if in the considered system the excitation energy migration occurs, then the rates (13) become insufficient, because then the process of energy transfer from component *i* to component *j* becomes multistep, and then in the process of transfer more or less important become correlations of positions of interacting molecules. Rates $\mathbb{k}_{ij}(t)$ do not take into account these correlations, hence expressions obtained using them for systems with migration must be considered approximate. A model that uses rates described by Eq. (13) is called a hopping model.

The model taking into account correlations of the positions of active molecules in the description of the transfer of excitation energy was introduced in [18] and developed in [19-21]. This model, referred to by the acronym SCDM (selfconsistent diagrammatic model), is thus considered by the authors to be inherently superior to the hopping model. This seems to be confirmed also by comparisons of the results predicted by this model with experimental data [38-41]. SCDM results generally describe the experimental data better than hopping model results. However, these works also show that also the results generated by the hopping model satisfactorily describe the experimental data, but mostly under the assumption of an additional excitation energy quenching mechanism caused by assuming a non-zero probability of its degradation during the transfer process. In addition, the results of the SCDM model are not exact but are the results of the so-called three-body approximation, the accuracy of which does not seem to be sufficiently studied. The disadvantage of the SCDM model is that it is extremely complicated and the necessary calculations are tedious. Currently, there is no extension of the SCDM to a multicomponent system. For these reasons, in this paper, the radiationless transfer of excitation energy between the individual components will be described only in the framework of the hopping model. To shorten and simplify the work, in our considerations we will not take into account the possibility of the aforementioned energy quenching accompanying the energy transfer process.

### 2.2 *Probability density functions and probabilities characterizing one-step deexcitation*

Suppose that a number of molecules of component $i$ were excited at time $t=0$. In general, the process of deexcitation of these molecules should be considered as a multi-step process during which the energy of excitation, before the final conversion to photon or heat, is repeatedly transferred both among molecules of component $i$ and molecules belonging to other components of the system. Let us now consider only the one-step deexcitation processes that immediately follow the moment of excitation. Of importance here are the distance-averaged $\mathcal{P}_{\mathrm{M}i}^{(1)}(t)$ probabilities of the excitation surviving on the $i$th component molecule from the moment of excitation at $t=0$ to any time $t \geq 0$. These probabilities can be understood as elements of the diagonal matrix $\mathcal{P}_{\mathrm{M}}^{(1)}(t)$

$$\mathcal{P}_{\mathrm{M}}^{(1)}(t) = \mathrm{diag}\left(\mathcal{P}_1^{(1)}(t), \ldots, \mathcal{P}_n^{(1)}(t)\right) \tag{15}$$

We will refer to the matrix (15) as the one-step MEE residence probability matrix. The temporal changes of the probability $\mathcal{P}_{\mathrm{M}i}^{(1)}(t)$ are described here by the differential equation

$$\frac{d\mathcal{P}_{\mathrm{M}i}^{(1)}(t)}{dt} = -\mathcal{P}_{\mathrm{M}i}^{(1)}(t)\left(\sum_{j=1}^{n} \mathbb{k}_{ij}(t) + k_{ri} + k_{nri}\right) \tag{16}$$

The solution of Eq. (16) with initial condition $\mathcal{P}_{\mathrm{M}i}^{(1)}(0) = 1$ is

$$\mathcal{P}_{\mathrm{M}i}^{(1)}(t) = \exp\left(-\int_0^t \sum_{j=1}^{n} \mathbb{k}_{ij}(t')\, dt' - \left(k_{ri} + k_{nri}\right)t\right) \tag{17}$$

In addition to probability $\mathcal{P}_{\mathrm{M}i}^{(1)}(t)$, the probability density $\rho_{\mathrm{M}i}^{(1)}(t)$ of occurrence of any of the events that deexcite the $i$th component molecule in one step is important for the description of NET in MCS. To determine $\rho_{\mathrm{M}i}^{(1)}(t)$ let us first note that using expression (17) we can determine the cumulative distribution function $\mathcal{F}_{\mathrm{M}i}^{(1)}(t)$ of such events

$$\mathcal{F}_{\mathrm{M}i}^{(1)}(t) \equiv 1 - \mathcal{P}_{\mathrm{M}i}^{(1)}(t) \tag{18}$$

The requested function $\rho_{\mathrm{M}i}^{(1)}(t)$ is the first derivative of this distribution function, by which, after considering (17) and (18), we obtain

$$\rho_{\mathrm{M}i}^{(1)}(t) = \mathcal{P}_{\mathrm{M}i}^{(1)}(t)\left(\sum_{j=1}^{n} \mathbb{k}_{ij}(t) + k_{ri} + k_{nri}\right) \tag{19}$$

It is easy to see that each of the $\rho_{\mathrm{M}i}^{(1)}(t)$ functions have the properties of the probability density function, and as such satisfies the normalization condition

$$\int_0^\infty \rho_{\mathrm{M}i}^{(1)}(t)\, dt = 1 \tag{20}$$

To better understand the properties of $\rho_{\mathrm{M}i}^{(1)}(t)$, let us rewrite Eq. (19) in the form

$$\rho_{\mathrm{M}i}^{(1)}(t) = \sum_{j=1}^{n} \Theta_{ij}^{(1)}(t) + \Phi_i^{(1)}(t) + Z_i^{(1)}(t) \tag{21}$$

In this equation, we can treat the terms $\Theta_{ij}^{(1)}(t)$ as elements of a square matrix $\Theta^{(1)}(t)$

$$\Theta^{(1)}(t) = \left[\Theta_{ij}^{(1)}(t)\right]_{n\times n} \tag{22}$$

The physical meaning of the function $\Theta_{ij}^{(1)}(t)$ is that the product $\Theta_{ij}^{(1)}(t)\, dt$ is equal to the probability of one-step NET from the $i$th component molecule to the $j$th component molecule in the interval $(t, t+dt)$. We will adopt the definition that the elements $\Theta_{ij}^{(1)}(t)$ are subnormalized probability density functions (SPDFs) describing the time evolution of the distance-averaged probability for NET between molecules of the same type $(i=j)$ and different types $(i \neq j)$. Justification for the term "subnormalized" used here can be found in the form of Eq. (31), where we always have $\theta_{ij}^{(1)} \leq 1$. Based on (19) and (21)), we can write

$$\Theta_{ij}^{(1)}(t) = \mathcal{P}_{\mathrm{M}i}^{(1)}(t)\, \mathbb{k}_{ij}(t) \tag{23}$$

In matrix notation, Eq. (23) takes the form of

$$\Theta^{(1)}(t) = \mathcal{P}_{\mathrm{M}}^{(1)}(t)\, \mathbb{k}(t) \tag{24}$$

where the $\mathcal{P}_{\mathrm{M}}^{(1)}(t)$ matrix is defined by Eq (15), and the $\mathbb{k}(t)$ matrix is given by Eqs (13) and (14). In Eq. (21), functions $\Phi_i^{(1)}(t)$ are SPDFs describing the time evolution of the average probability for spontaneous conversion of the excitation energy, actually residing on the $i$th kind molecule, to photon. The physical meaning of function $\Phi_i^{(1)}(t)$ is that the product $\Phi_i^{(1)}(t)\, dt$ is equal to the average probability of quantum of light emission from the $i$th component molecule in the interval $(t, t+dt)$, assuming that this molecule was excited at time $t=0$. We will consider these functions as elements of the diagonal matrix $\Phi^{(1)}(t)$

$$\Phi^{(1)}(t) = \mathrm{diag}\left(\Phi_1^{(1)}(t), \ldots, \Phi_n^{(1)}(t)\right) \tag{25}$$

whereby, based on (19) and (21), we assume that

$$\Phi^{(1)}(t) = \mathcal{P}_{\mathrm{M}}^{(1)}(t)\, k_{\mathrm{r}} \tag{26}$$

where the $k_{\mathrm{r}}$ matrix is defined by Eq. (11). Finally, the functions $Z_i^{(1)}(t)$ in Eq. (21) are SPDFs describing the time evolution of the average probability for the excitation energy internal conversion into heat. The physical meaning of the function $Z_i^{(1)}(t)$ is that the product $Z_i^{(1)}(t)\, dt$ is equal to the average probability of degradation of MEE to heat on the $i$th component in the time interval $(t, t+dt)$, assuming that this molecule was excited at time $t=0$. By analogy with (25) and (26), we will assume that

$$Z^{(1)}(t) = \mathrm{diag}\left(Z_1^{(1)}(t), \ldots, Z_n^{(1)}(t)\right) \tag{27}$$

$$Z^{(1)}(t) = \mathcal{P}_{\mathrm{M}}^{(1)}(t)\, k_{\mathrm{nr}} \tag{28}$$

where the $k_{\mathrm{nr}}$ matrix is defined by Eq. (12). After taking into account (12) and (26), one can also write

$$Z^{(1)}(t) = \Phi^{(1)}(t)\left(I_n - \phi_0\right)\phi_0^{-1} \tag{29}$$

Note that it follows from Eqs. (20) and (21) that the functions $\Theta_{ij}^{(1)}(t)$, $\Phi_i^{(1)}(t)$, and $Z_i^{(1)}(t)$ are normalized in the sense given below

$$\int_0^\infty \left(\sum_{j=1}^{n} \Theta_{ij}^{(1)}(t) + \Phi_i^{(1)}(t) + Z_i^{(1)}(t)\right) dt = 1 \tag{30}$$

### 2.2.1 One-step NET kinetics under steady-state conditions

Under steady-state conditions, both the intensity of the excitation light and the intensity of the fluorescence emitted by the MCS is constant over time. Then the processes of excitation energy transfer and conversion cease to depend directly on shape of the time courses of $\Theta_{ij}^{(1)}(t)$, $\Phi_i^{(1)}(t)$, and $Z_i^{(1)}(t)$ functions contained in the $\Theta^{(1)}(t)$, $\Phi^{(1)}(t)$, and $Z^{(1)}(t)$ matrices, respectively, but depend on $\theta_{ij}^{(1)}$, $\phi_i^{(1)}$, and $\zeta_i^{(1)}$ probabilities which are the integrals of $\Theta_{ij}^{(1)}(t)$, $\Phi_i^{(1)}(t)$, and $Z_i^{(1)}(t)$ functions taken over time from zero to infinity. These probabilities refer to events that may occur after the arrival of MEE on the component *i* molecule. They can be represented as matrices $\theta^{(1)}$, $\phi^{(1)}$, and $\zeta^{(1)}$ satisfying the conditions

$$\theta^{(1)} = \left[\theta_{ij}^{(1)}\right]_{n\times n} = \int_0^\infty \Theta^{(1)}(t)\,dt \tag{31}$$

$$\phi^{(1)} = \operatorname{diag}\left(\phi_1^{(1)},\ldots,\phi_n^{(1)}\right) = \int_0^\infty \Phi^{(1)}(t)\,dt \tag{32}$$

$$\begin{aligned}\zeta^{(1)} &= \operatorname{diag}\left(\zeta_1^{(1)},\ldots,\zeta_n^{(1)}\right) = \int_0^\infty Z^{(1)}(t)\,dt \\ &= \phi^{(1)}\left(I_n - \phi_0\right)\phi_0^{-1}\end{aligned} \tag{33}$$

where $\theta_{ij}^{(1)}$ denotes the probability of one-step transferring of MEE to the *j*th component molecule, $\phi_i^{(1)}$ denotes the probability of emitting MEE in the form of photon, and $\zeta_i^{(1)}$ denotes the probability of converting MEE into heat. It follows from Eqs. (30)-(33) that

$$\sum_{j=1}^{n}\theta_{ij}^{(1)} + \phi_i^{(1)} + \zeta_i^{(1)} = 1 \tag{34}$$

In the next section we find expressions to calculate the values of $\Theta_{ij}^{(1)}(t)$, $\Phi_i^{(1)}(t)$, $Z_i^{(1)}(t)$, $\theta_{ij}^{(1)}$, $\phi_i^{(1)}$, and $\zeta_i^{(1)}$ parameters in the presence of Förster resonance energy transfer (FRET).

### 2.2.2 Distance-averaged one-step parameters in the case of FRET

If the NET process in the considered MCS has a dipole-dipole character, then the rate $k_{ij}(r)$ in Eq. (13) can be described by the equation [42]

$$k_{ij}(r) = \frac{1}{\tau_{0i}}\left(\frac{R_{0ij}}{r}\right)^6 \tag{35}$$

where *r* is the distance between interacting molecules, and $R_{0ij}$ is the Förster radius [34]. Insertion of (35) into (13) yields

$$\mathbb{K}_{ij}^{(1)}(t) = \frac{\gamma_{ij}}{\sqrt{\tau_{0i}\,t}}\operatorname{erf}\left[\frac{R_{0ij}^3}{a_{ij}^3}\sqrt{\frac{t}{\tau_{0i}}}\right] \tag{36}$$

where

$$\gamma_{ij} = \frac{2}{3}\pi^{3/2}R_{0ij}^3\,c_j \tag{37}$$

Expression (36) simplifies significantly if one assumes that the dimensions of the interacting molecules are negligibly small, or alternatively, that their minimum mutual distance $a_{ij}$ is equal to zero. Then we obtain

$$\mathbb{K}_{ij}^{(1)}(t) = \frac{\gamma_{ij}}{\sqrt{\tau_{0i}\,t}} \tag{38}$$

Equation (38) corresponds to Eq. (8.3) in [43]. Because of its uncomplicated structure and good agreement with experimental data the expression (38) was in its more or less explicit form used by many authors [6,11,17,32,35,36]. In the present work, we will also use this expression. Inserting (38) into (17) leads to

$$\mathcal{P}_i^{(1)}(t) = g(t;\Gamma_i,\tau_{0i}) \tag{39}$$

where

$$\Gamma_i = \sum_{j=1}^{n}\gamma_{ij} \tag{40}$$

and

$$g(t,\Gamma,\tau) = \exp\left(-2\Gamma\sqrt{\frac{t}{\tau}} - \frac{t}{\tau}\right) \tag{41}$$

The combination of (38), (39), and (23) allows us to find $\Theta_{ij}^{(1)}(t)$

$$\Theta_{ij}^{(1)}(t) = \frac{\gamma_{ij}}{\sqrt{\tau_{0i}\,t}}\,g(t;\Gamma_i,\tau_{0i}) \tag{42}$$

After inserting (39) sequentially into (26) and (28), we find $\Phi_i^{(1)}(t)$ and $Z_i^{(1)}(t)$.

$$\Phi_i^{(1)}(t) = \frac{\phi_{0i}}{\tau_{0i}}\,g(t;\Gamma_i,\tau_{0i}) \tag{43}$$

$$Z_i^{(1)}(t) = \frac{1-\phi_{0i}}{\tau_{0i}}\,g(t;\Gamma_i,\tau_{0i}) \tag{44}$$

Averaged over distances, the probability $\theta_{ij}^{(1)}$ of a one-step transfer can be calculated from (31), (23), (38), (39), and (41)

$$\theta_{ij}^{(1)} = \frac{\gamma_{ij}}{\Gamma_i}\,f_{\mathrm{F}}\left(\Gamma_i\right) \tag{45}$$

where

$$f_{\mathrm{F}}(x) = \sqrt{\pi}\,x\exp\left(x^2\right)\left[1-\operatorname{erf}(x)\right] \tag{46}$$

Distance averaged probability of one-step photon emission can be calculated from (32), (26), (39), and (41). This yields

$$\phi_i^{(1)} = \phi_{0i}\left[1 - f_{\mathrm{F}}\left(\Gamma_i\right)\right] \tag{47}$$

Eq. (47) have been first obtained by Förster [6]. Given $\phi_i^{(1)}$, the distance-averaged probability of one-step excitation conversion into heat, $\zeta_i^{(1)}$, can be calculated from Eq. (33)

$$\zeta_i^{(1)} = \left(1-\phi_{0i}\right)\left[1 - f_{\mathrm{F}}\left(\Gamma_i\right)\right] \tag{48}$$

## 3 Multistep NET kinetics in MCS. Finding the values of individual elements of matrices $\Phi(t)$ and $\phi$

In this section, assuming the hopping mechanism of NET, we will discuss how to find expressions to calculate the values of the elements of the matrices $\Phi(t)$ and $\phi$ under conditions of multi-stage, forward and reverse energy transfer between all components of the considered MCS.

### 3.1 Introductory remarks

At the outset, it should be acknowledged that the expressions to calculate the values of the elements of the $\Phi(t)$ and $\phi$ matrices can be found by slightly modifying (see also Section 3.10) the method described in the paper [32]. This method amounts to solving an appropriate system of algebraic equations. In this work, however, we will use a simpler, and in our opinion more intuitive, method using the Markov chain technique for the same purpose.

The Markov chain technique is based on the fact that the internal changes of the described system are modeled in such a way that the current state of the chain object wanders among a finite number of possible states, according to the probability distribution functions assigned to each of these states. In the case of MCS, we determine that the Markov chain object is MEE. The time evolution of MEE takes the form of a chain of individual stages or steps. From Eqs. (42)-(48), it can be seen that, under our conditions, the probabilities of MEE transition between states do not depend on the value of the $\mu$ index, which means that the Markov chains describing MEE transfer in MCS are homogeneous. A closer analysis of issues related to Markov chains suggests that methods operating in discrete state space may be useful for describing MCS fluorescence, in particular: 1) the method known as discrete time Markov chains (DTMC), 2) the method known as continuous time Markov chains (CTMC), and 3) the method resulting from theoretical considerations classified as renewal theory (RT).

The DTMC method operates with time-independent probabilities of transitions between states, which suggests that it may be appropriate for describing the fluorescence properties of a given MCS only under steady-state conditions, when the averaged probabilities of one-step processes are constant in time. Taking into account the fact that the averaged transition probabilities (45)-(48) remain constant over time, we can claim that the Markov chain formed with these probabilities will be a stationary stochastic process. Unfortunately, the name DTMC is misleading, since time as such does not appear in the DTMC model. Within DTMC, it is customary, without ensuring the necessary accuracy, to assign the term "time" to the values of $\mu$ index specified in Eq. (4). However, this index is not time, if only because it does not have the dimension of time. The real time $t$ can possibly be introduced into the DTMC, by assuming the relationship holds

$$t_\mu = \mu h \tag{49}$$

where $h > 0$ is the chosen time segment. Under this assumption, the values of $t_\mu$ indicate the moments of time at which the MEE transition from one state to another occurs. However, when a given Markov chain involves a steady state, the $t_\mu$ moments have no experimentally determinable counterparts.

Under pulse excitation of MCS fluorescence, the averaged probabilities of one-step processes depend on time, for example, as shown by expressions (42)-(44). In the CTMC method, the probabilities of one-step processes are treated as time-dependent, however, these dependencies must be exponential [44,45], which is not fulfilled in Eqs. (42)-(44). To solve this problem, in this work we introduce a new type of Markov chains, which we call time-resolved Markov chains (TRMC). In a sense, the TRMC method is an extension of the DTMC method to the case where the probabilities of transitions between process states depend on time in any way. In addition, the TRMC method, unlike the CTMC method, allows consideration of excitation energy migration processes among molecules of the same component, whereas under CTMC, transitions within the same state are not allowed [46].

It seems that among the theoretical studies known to date, directly or indirectly related to use of Markov chains to describe MCS fluorescence, the most useful may be the methods resulting from considerations carried out within RT [47-51]. An important property of RT is that, unlike CTMC, the dependencies of the probability densities of single-step MEE transfers on time can be arbitrary, not just exponential. The application of RT to the description of MCS fluorescence requires a series of adjustments similar to those we perform in this work when discussing the TRMC method. Therefore, the possibility of using RT instead of TRMC in our considerations is discussed in Section 3.9, after clarifying our TRMC method in Section 3.5.

### 3.2 Kinds of MEE states and transitions in MCS

The time evolution of a single MEE in MCS involves its random wandering among fluorescing molecules, which in each case ends with either the emission of a photon of fluorescence or the conversion of a given MEE to heat. We will call the different stages of this wandering the MEE states. MEE transitions from one state to another occur with a certain probability and occur at random but statistically distributed moments of time. Thus, we can understand MEE wandering in MCS as a kind of stochastic process governed by two random variables, one of which determines the type of state to which the MEE transitions, and the other determines the moment of time at which this transition occurs. In the case of MCS, we distinguish three types of states that can characterize an MEE:

a. ***States in which the MEE is localized to one of the fluorescent molecules.*** We will assume that the number of such states is equal to the number of $n$ fluorescent components present in the system under consideration.
b. ***States in which MEE is immediately converted to a photon.*** It seems obvious that the total number of states of this type assigned to a given system cannot be greater than the number $n$ of components. The specific number of such states assumed depends on how much information concerning the dynamics of fluorescence decay we want to acquire. Typically, from our considerations we want to extract information about the time distribution of photons emitted by each individual component, hence we will generally assume that the number of such states is equal to the number $n$ of components in the system.
c. ***States in which MEE is immediately converted to heat.*** The total number of states in this group, as in (b), depends not only on the number of components but also on how much information regarding the radiationless deactivation of the excitation we want to obtain. Among the various variants of the organization of this group, it is possible, for example, to represent all radiationless relaxation acts as transitions to one common state.

Note that type (a) MEE states can be produced either by direct excitation with external light or by MEE transitions between type (a) states, while type (b) and (c) states can only be produced by MEE transitions from type (a) states.

Possible MEE transitions between individual states are represented in the form of a chain called a Markov chain. A given Markov chain can contain any number of states. In DTMC terminology, states of type (a) should be classified as transient states, because there is always a non-zero probability of transition from states (a) to states (b) or (c). If transitions from a given state of type (a) to the same state are also possible, then this state also acquires some of the characteristics of an absorbing state. Partial properties of an absorbing state can be assigned to states associated with the presence of MEE on molecules for which it is possible to transfer MEE to molecules of the same type. States (b) and (c) behave similarly to absorbing states, but these states cannot be strictly classified as absorbing states because, once they are reached, any events involving MEE in a given chain cease to occur. This is because for states of type (b) or (c), the probability of transition to all states in a given chain is assumed to be zero. Therefore, we classify such states as a new type of states, which we call terminating states. The properties of a terminating state differ from those of an absorbing state in the sense that, once the absorbing state has been reached, further transitions of the chain object are possible, but only to the same absorbing state, whereas once the terminating state has been reached, no transitions of the chain object between states are possible. In our MCS fluorescence model, absorbing states do not occur at all. Terminating states, on the other hand, are used to describe the state of MEE after its conversion to a photon or heat. One of the advantages of using the terminating state instead of the absorbing state is that, for example, the quantum efficiency of fluorescence can be directly interpreted as the number of visits to the terminating state, which cannot be done with the absorbing state. A more detailed analysis of the properties of terminating, absorbing, and partially absorbing states is carried out in Section 4.3.

Each one-step transition of a given MEE from one state to another is classified as an elementary event. We will assume that we know the time distributions of these events as given by known time-dependent SPDFs. In finding expressions describing the elements of the $\varPhi(t)$ and $\phi$ matrices, we will consider only those elementary events involving MEE that are caused by NET and by the spontaneous processes of emission and internal conversion. These events are

i. ***Nonradiative transfer of the MEE from the i*th *component molecule to the j*th *component molecule.*** The probability of one-step transferring the excitation energy between these molecules is equal to $\theta_{ij}^{(1)}$ calculated from Eqs. (31) and (45). The SPDFs describing the time distribution of these events after the act of excitation are given by $\varTheta_{ij}^{(1)}(t)$ functions defined by Eqs. (23) and (42). As a result of such transfers a given MEE goes from state $i$ to state $j$. If $j = i$, that is, when the energy transfer occurs between molecules of the same component, then we consider that after the act of transfer the given MEE remains in state $i$.

ii. ***Spontaneous conversion of the MEE to photon (emission of fluorescence).*** The probability for one-step photon emission is given by $\phi_i^{(1)}$ expressed by Eqs. (32) and (47). The SPDFs describing the time distribution of such events after excitation are given by functions $\varPhi_i^{(1)}(t)$ expressed by Eqs. (26) and (43).

iii. ***Internal conversion of the MEE to heat (internal de-excitation).*** The probability for one-step internal de-excitation is given by $\zeta_i^{(1)}$ expressed by Eqs. (33) and (48). The SPDFs describing the time distribution of such events after excitation are given by functions $Z_i^{(1)}(t)$ expressed by Eqs. (28) and (44).

iv. ***Events that could potentially occur after MEE reaches the terminating state.*** Since in the terminating state, MEE loses the ability to transfer to any state in a given chain, we assume that all SPDFs describing MEE transitions from each terminating state are equal to zero.

In our considerations, the total number of states characterizing MEE in a given MCS and taken into account in the Markov chain under consideration is equal to $\nu = n + m$, where $n$ is the number of components and $m$ is the assumed number of terminating states.

### 3.3 Probability distributions of one-step interstate transitions. Markov property

Given Eqs. (22)-(29), one can conclude that any type of one-step interstate MEE transition in MCS is governed by a probability distribution of bivariate random variable $(J,T)$ [52-54]. The random variable $J$ indicates the number of the state the MEE will be in during the next step. It is a discrete variable expressed by $j$ values from the set $\mathcal{S} = \{1,2,\ldots,\nu\}$. The random variable $T$ determines the moments of MEE transition between states. It is a continuous variable expressed by values of $t$ from the set $\mathcal{T} = \mathbb{R}_+^0$. The values of both variables are determined by assuming that at time $t = 0$ MEE is located on one of the component $i$ molecules. Under such conditions, for each state $i \in \mathcal{S}$ we have a separate joint probability mass-density function $\mathbb{P}_{iJT}^{(1)}(j,t)$ having the property that the product of $\mathbb{P}_{iJT}^{(1)}(j,t)\,dt$ is equal to the probability of one-step transition of MEE from state $i$ to state $j$ in the time interval $(t, t+dt)$. For $t < 0$, the $\mathbb{P}_{iJT}^{(1)}(j,t)$ functions are zero, while for $t \geq 0$, the exact form of each of these functions depends on the mechanism adopted for the transfer of MEE between states, the number of terminating states assumed, and the individual properties of these states. For simplicity of notation, we will write the function $\mathbb{P}_{iJT}^{(1)}(j,t)$ as $\mathbb{P}_{ij}^{(1)}(t)$. For example, if we assign to each component of the solution one terminating state corresponding to photon emission and one terminating state corresponding to the conversion of MEE to heat, then $m = 2n$, $\nu = 3n$. Under these conditions, for $t \geq 0$, the functions $\mathbb{P}_{ij}^{(1)}(t)$ can be given by the expression

$$\mathbb{P}_{ij}^{(1)}(t)=\begin{cases}\left.\begin{aligned}&\Theta_{ij}^{(1)}(t), &&\text{if } (1\le j\le n\\ &\Phi_i^{(1)}(t)\,\delta_{i,(j-n)}, &&\text{if } (n<j\le 2n\\ &Z_i^{(1)}(t)\,\delta_{i,(j-2n)}, &&\text{if } (2n<j\le \nu\end{aligned}\right\}\text{and } 1\le i\le n)\\ 0, \qquad\qquad\qquad\quad \text{if } n<i\le\nu\end{cases} \tag{50}$$

where $\delta_{i,j}$ is the Kronecker delta. We constructed the functions $\mathbb{P}_{ij}^{(1)}(t)$ in such a way that with respect to the random variable $J$ they are probability mass functions, and with respect to the random variable $T$ they are probability density functions. The stochastic process defined in this work as TRMC is a sequence of values of a bivariate random variable $(J,T)$, which can be written as

$$\left\{\left(j_\mu, t_\mu\right); \mu = 2,3\ldots\right\} \tag{51}$$

Here $j_\mu$ is the state number in which the MEE is in during the $\mu$th step of the process, and the value of $t_\mu$ determines the time at which the MEE transitions to that state. For step $\mu=1$, we take $t_1=0$, and the value of $j_1$ specifies the number of the component excited by the external light beam. It can be assumed that the process (51) is Markovian if for $j_{\mu+1}\in\mathcal{S}$, $t_{\mu+1}\in\mathcal{T}$, and $\mu\in\mathcal{M}$, there occurs

$$\begin{aligned}\Pr\Big\{\left(J_{\mu+1}=j_{\mu+1}, T_{\mu+1}=t_{\mu+1}\right)\Big|\left(J_2=j_2, T_2=t_2\right),&\\ \left(J_3=j_3, T_3=t_3\right),\cdots,\left(J_\mu=j_\mu, T_\mu=t_\mu\right)\Big\}&\\ =\Pr\Big\{\left(J_{\mu+1}=j_{\mu+1}, T_{\mu+1}=t_{\mu+1}\right)\Big|\left(J_\mu=j_\mu, T_\mu=t_\mu\right)\Big\}&\end{aligned} \tag{52}$$

for any set of states $j_2,\cdots,j_\mu$ in $\mathcal{S}$ and any set of times $t_2,\cdots,t_\mu$ in $\mathcal{T}$. The denotation $\mathbf{Pr}$ in (52) stands for probability measure. Non-Markovian process, on the other hand, exhibits memory effects, meaning that the future state depends not only on the present state but also on past states and the specific path the process has taken to reach its current state. For TRMC, we assume that the one-step distributions of the random variable $(J,T)$ taken for different values of $\mu$ are mutually independent. At any step $\mu>0$, based only on the parameters of the $\mathbb{P}_{ij}^{(1)}(t)$ function, we are able to calculate, using e.g. the Gillespie algorithm [55], the values of the random variables $j_{\mu+1}$ and $t_{\mu+1}$ for the step $\mu+1$. Therefore, we believe that TRMC fulfills the condition (52) and that it should be qualified as Markovian even if the associated one-step time distributions of the $(J,T)$ variable are not memoryless.

There are two marginal distributions associated with each of the distributions defined by Eq. (50). The marginal one-step distributions $R_i^{(1)}(t)$ of the random variable $T$ are obtained by summing (50) over all values of $j$. For $t<0$ the $R_i^{(1)}(t)$ functions are zero, while for $t\ge 0$, this gives

$$R_i^{(1)}(t)=\sum_{j=1}^{\nu}\mathbb{P}_{ij}^{(1)}(t)=\begin{cases}\rho_i^{(1)}(t), & \text{if } 1\le i\le n\\ 0, & \text{if } n<i\le\nu\end{cases} \tag{53}$$

where $\rho_i^{(1)}(t)$ are defined by Eq. (19). The function $R_i^{(1)}(t)$ represents the temporal probability density distribution of any event in state $i$. The marginal distributions $P_{ij}^{(1)}$ of the random variable $J$ are obtained by integrating (50) over all values of $t$

$$P_{ij}^{(1)}=\int_0^\infty \mathbb{P}_{ij}^{(1)}(t)\,dt \tag{54}$$

This gives

$$P_{ij}^{(1)}=\begin{cases}\left.\begin{aligned}&\theta_{ij}^{(1)}, &&\text{if } 1\le j\le n\\ &\phi_i^{(1)}\,\delta_{i,(j-n)}, &&\text{if } n<j\le 2n\\ &\zeta_i^{(1)}\,\delta_{i,(j-2n)}, &&\text{if } 2n<j\le \nu\end{aligned}\right\}\text{and } 1\le i\le n\\ 0, \qquad\qquad\qquad \text{if } n<i\le\nu\end{cases} \tag{55}$$

$P_{ij}^{(1)}$ distributions are univariate discrete probability distributions of the random variable $J$. As can be seen from Eq. (54), the $P_{ij}^{(1)}$ distributions describe the summary probabilities of transitions between states without indicating the moments at which these transitions occur. Therefore, we called the stochastic process associated with these distributions a time-unspecified Markov chain (TUMC). Time here ceases to be a random variable, hence TUMC involves only one random variable $J$. This process can be written as

$$\left\{j_\mu; \mu = 2,3,\ldots\right\} \tag{56}$$

TUMC is Markovian if it occurs

$$\begin{aligned}&\Pr\Big\{\left(J_{\mu+1}=j_{\mu+1}\right)\Big|\left(J_2=j_2\right),\left(J_3=j_3\right),\cdots,\left(J_\mu=j_\mu\right)\Big\}\\ &\quad=\Pr\Big\{\left(J_{\mu+1}=j_{\mu+1}\right)\Big|\left(J_\mu=j_\mu\right)\Big\}\end{aligned} \tag{57}$$

Because the probability distributions underlying TUMC satisfy relation (54), this process can be applied to steady-state NET analysis. Note that the process referred to here as TUMC is equivalent to the process known as DTMC. However, since it is our understanding that DTMC does not refer to time, we propose to rename it TUMC.

### *3.4 The one-step transition matrices*

The basis of the TRMC method is to collect all SPDFs characterizing one-step EEM transitions in a given MCS as elements of an $(\nu\times\nu)$-dimensional matrix $\mathbb{P}^{(1)}(t)$, which we will call the one-step transition SPDF matrix. We prefer to write these matrices in the so-called canonical form. In this form, the first $n$ rows of the $\mathbb{P}^{(1)}(t)$ matrix refer to transient states. In these states, MEE is located on one of the molecules of one of the $n$ components of our system, and the row number assigned to a given state is the number of that component. For $1\le i\le n$, the elements of row $i$ in the matrix $\mathbb{P}^{(1)}(t)$ sequentially are the SPDFs $\Theta_{ij}^{(1)}(t)$, $\Phi_{ij}^{(1)}(t)$, and $Z_{ij}^{(1)}(t)$. The remaining $m$ rows of the matrix $\mathbb{P}^{(1)}(t)$ refer to states in which the MEE loses the ability to transition to any of the states of a given chain. Therefore, we assume that all elements of the $\mathbb{P}^{(1)}(t)$ matrix located in the rows corresponding to such states are zeros. In the fluorescent systems, states of this type characterize the MEE after its conversion to photon or heat and we classify them as terminating states.

One of the simplest one-step transition SPDF matrices characterizing MCS can be constructed when the number of solution components is limited to $n=2$. The time-dependent fluorescence of such a system can be analyzed using the six-state Markov chain shown in Fig. 1. In this figure, the red circles correspond to transient states, and the blue circles and gray circles correspond to terminating states. In states 1 and 2, MEE resides on molecules belonging to component 1 or 2, respectively. In states 3 and 4, MEE is immediately converted to photon which is observed as fluorescence emitted by component 1 or 2, respectively.

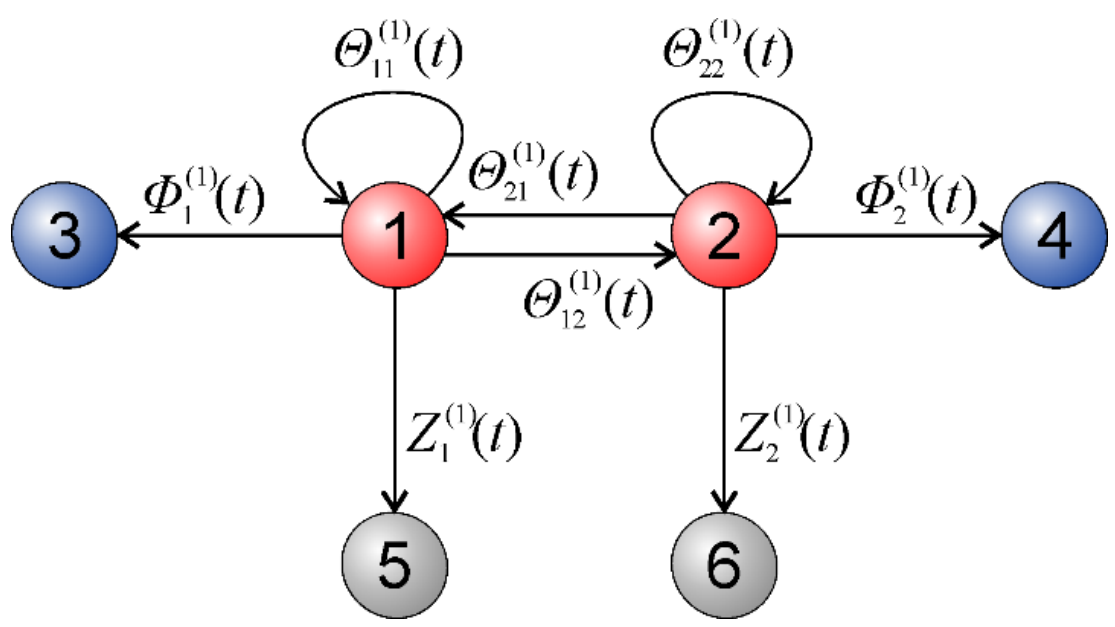

**Figure 1.** Six-state Markov process diagram showing states, elementary events and their SPDFs in a two-component system. The red circles illustrate the states of transient localization of MEE on the molecules of component 1 or 2, respectively. The blue circles and gray circles illustrate the terminating states in which MEE is immediately converted into either photons or heat, respectively.

When state 5 or 6 is reached, MEE is internally converted to heat. In this approach, the number of transient states is given by $n=2$, the number of terminating states is given by $m=4$, making $\mathbb{P}^{(1)}(t)$ a $6\times 6$ matrix of the form

$$\mathbb{P}^{(1)}(t) = \begin{bmatrix} \Theta_{11}^{(1)}(t) & \Theta_{12}^{(1)}(t) & \Phi_{1}^{(1)}(t) & 0 & Z_{1}^{(1)}(t) & 0 \\ \Theta_{21}^{(1)}(t) & \Theta_{22}^{(1)}(t) & 0 & \Phi_{2}^{(1)}(t) & 0 & Z_{2}^{(1)}(t) \\ 0 & 0 & 0 & 0 & 0 & 0 \\ 0 & 0 & 0 & 0 & 0 & 0 \\ 0 & 0 & 0 & 0 & 0 & 0 \\ 0 & 0 & 0 & 0 & 0 & 0 \end{bmatrix} \quad (58)$$

The row numbers of this matrix correspond to the state numbers illustrated in Fig. 1. As elements of the first row of matrix (58), we have SPDFs of one-step events possible when MEE resides on molecules of component 1. The fourth element of this row is zero because we assume that state 4 is directly reachable only from state 2. Similarly, the sixth element of this row is zero because we assume that state 6 is also directly reachable only from state 2. The second row of matrix $\mathbb{P}^{(1)}(t)$, which deals with possible events when MEE resides on molecules of component 2, is constructed analogously. Rows 3-6 of matrix (58) contain SPDFs of MEE transitions that, according to our model, could occur from terminating states to all possible states. In our model, all these SPDFs are zero. For further considerations, it is advantageous to divide the $\mathbb{P}^{(1)}(t)$ matrix into blocks according to the scheme

$$\mathbb{P}^{(1)}(t) = \begin{bmatrix} \Theta^{(1)}(t) & \Psi^{(1)}(t) \\ O_{m\times n} & O_{m\times m} \end{bmatrix} \quad (59)$$

Here $\Theta^{(1)}(t)$ is an $n\times n$ submatrix describing transitions between transient states, $\Psi^{(1)}(t)$ is an $n\times m$ submatrix describing the excitation transitions from the transient to terminating states, and $O_{m\times n}$ and $O_{m\times m}$ are the $m\times n$ and an $m\times m$ zero submatrices, respectively. SPDFs that are elements of the $\Theta^{(1)}(t)$ submatrix are expressed by $\Theta_{ij}^{(1)}(t)$ functions defined by Eq. (23), and in the case where the transfer mechanism is FRET, they can be expressed by functions given by Eq. (42). The submatrix $\Psi^{(1)}(t)$ consists of two subsubmatrices of size $n\times n$, $\Phi_{i}^{(1)}(t)$ and $Z_{i}^{(1)}(t)$, according to the scheme

$$\Psi^{(1)}(t) = \begin{bmatrix} \Phi^{(1)}(t) & Z^{(1)}(t) \end{bmatrix} \quad (60)$$

All SPDFs in $\Psi^{(1)}(t)$ describe the transitions of MEE from a particular component distinguished by the row number $i$ to the terminating states distinguished by the column number $j=n+l$, where $l$ is a column number in submatrix $\Psi^{(1)}(t)$. The diagonal subsubmatrix $\Phi^{(1)}(t)$ has the form given by Eq. (25). Its $\Phi_{i}^{(1)}(t)$ elements are defined by Eq. (26), and in the case of FRET can be functions given by Eq. (43). The diagonal subsubmatrix $Z^{(1)}(t)$ has the form given by Eq. (27), and its elements $Z_{i}^{(1)}(t)$ are defined by Eq. (28), and in the case of FRET can be functions given by Eq. (44). The elements of the transient rows of the matrix $\mathbb{P}^{(1)}(t)$ are not normalized in such a way that their sum in each row is equal to unity, but are normalized in such a way that the integral of the sum of the elements in a given row, taken from zero to infinity, is equal to unity

$$\int_0^\infty \sum_{j=1}^{\nu} \mathbb{P}_{ij}^{(1)}(t)\,dt = \sum_{j=1}^{\nu} P_{ij}^{(1)} = 1, \qquad 1 \le i \le n \quad (61)$$

The latter property follows from Eq. (30) and (54).

In the case when MCS fluorescence is excited with continuous light, then the TUMC formalism can be used instead of the TRMC formalism to describe MEE transfer. Then, according to (54) and (55), the probabilities of MEE transitions between possible states of the system shown in Fig. 1 are described by a time-independent $P^{(1)}$ matrix of the form

$$P^{(1)} = \begin{bmatrix} \theta_{11}^{(1)} & \theta_{12}^{(1)} & \phi_{1}^{(1)} & 0 & \zeta_{1}^{(1)} & 0 \\ \theta_{21}^{(1)} & \theta_{22}^{(1)} & 0 & \phi_{2}^{(1)} & 0 & \zeta_{2}^{(1)} \\ 0 & 0 & 0 & 0 & 0 & 0 \\ 0 & 0 & 0 & 0 & 0 & 0 \\ 0 & 0 & 0 & 0 & 0 & 0 \\ 0 & 0 & 0 & 0 & 0 & 0 \end{bmatrix} \quad (62)$$

The $P^{(1)}$ matrix is called the one-step transition probability matrix. As with the $\mathbb{P}^{(1)}(t)$ matrix, it is convenient to divide the $P^{(1)}$ matrix into blocks according to the scheme

$$P^{(1)} = \begin{bmatrix} \theta^{(1)} & \psi^{(1)} \\ O_{m\times n} & O_{m\times m} \end{bmatrix} \quad (63)$$

Here $\theta^{(1)}$ is an $n\times n$ submatrix describing transitions between transient states, $\psi^{(1)}$ is an $n\times m$ submatrix describing the excitation transitions from the transient to terminating states, while $O_{m\times n}$ and $O_{m\times m}$ denote the $m\times n$ and an $m\times m$ zero matrices, respectively. In our case, the elements of the $\theta^{(1)}$ submatrix are $\theta_{ij}^{(1)}$ probabilities given by Eq. (45). As with $\Psi^{(1)}(t)$ submatrix, the $\psi^{(1)}$ submatrix can be represented as a composite of two subsubmatrices of size $n\times n$, $\phi^{(1)}$ and $\zeta^{(1)}$

$$\psi^{(1)} = \left[\phi^{(1)} \quad \zeta^{(1)}\right] \tag{64}$$

The $\phi^{(1)}$ subsubmatrix is the same as matrix (32) and contains the $\phi_i^{(1)}$ probabilities for one-step light quantum emission by the molecules of the individual components of the system calculated, for example, from Eq. (47). The $\zeta^{(1)}$ subsubmatrix is the same as matrix (33) and contains the $\zeta_i^{(1)}$ probabilities of one-step internal conversion of the excitation energy to heat, calculated, for example, from Eq. (48). As can be seen from Eq. (61) , the sum of all elements in each of the first *n* rows of matrix $P^{(1)}$ is equal to unity.

### *3.4.1 The problem of stochasticity of the transition matrices*

One of the basic assumptions in considerations concerning Markov chains is the stochasticity of the transition matrices. With our notation, this means that to be stochastic, one-step transition matrices should be nonnegative, and the sum of the elements in each of their rows should be equal to one [56-61]. The above assumption can also be formulated in such a way that the individual rows of the transition matrix are probability vectors [61-64]. Such properties are overtly observed in DTMC and less overtly in CTMC, where so-called rate matrices become more important than transition matrices.

In this work, the rows of the transition matrices corresponding to the terminating states are completely filled with zeros, and in the case of TRMC, the elements of these matrices are not probabilities, but probability densities. Therefore, the transition matrices discussed in this work cannot be qualified as stochastic. From the point of view of our considerations, the problem of stochasticity of the transition matrix is not essential. In the TUMC framework, the most important thing is that the elements of the transition matrix have the meaning of the probabilities of transition from one state to another. Similarly, under TRMC, what is most important is that the elements of the transition matrix have the meaning of SPDFs describing the probability densities of transitions between states. In both cases, the properties of the sum of these elements in a given row are of secondary significance.

### *3.4.2 Probability of reaching a given state in multiple steps*

The probability distributions $P_{ij}^{(1)}$ and $\mathbb{P}_{ij}^{(1)}(t)$ describe MEE transitions from state *i* to state *j* in one step. From the point of view of describing the fluorescence intensities of the individual components of MCS, it is important to know the probabilities of reaching the selected states in multiple steps. In the case of TUMC, we assume that the probability $P_{ij}^{(\mu)}$ of MEE reaching state *j* in *μ*th step, given that $\mu > 1$ and that in step $\mu = 1$ MEE was in state *i,* can be calculated in the same way as for DTMC [45,46,57,60,64]. In our notation, this means

$$P_{ij}^{(\mu)} = \sum_{q=1}^{\nu} P_{iq}^{(\mu-1)} P_{qj}^{(1)} \tag{65}$$

In the case of TRMC, when we use the bivariate probability distributions $\mathbb{P}_{ij}^{(\mu)}(t)$ of reaching state *j* from state *i* at time *t* as a result of the *μ*-step transfer of excitation energy, relation (65) requires modification. This modification consists in replacing simple products of probabilities by convolutions of the corresponding SPDFs. This is because in this case the random variable $T^{(\mu)}$ is the sum of the independent random variables $T^{(\mu-1)}$ and $T^{(1)}$. Thus, we will assume that the bivariate probability distributions $\mathbb{P}_{ij}^{(\mu)}(t)$ of reaching state *j* from state *i* at time *t*, as a result of *μ*-step transfer of excitation energy, are given by the expression

$$\mathbb{P}_{ij}^{(\mu)}(t) = \sum_{q=1}^{\nu} \left(\mathbb{P}_{iq}^{(\mu-1)} * \mathbb{P}_{qj}^{(1)}\right)(t) \tag{66}$$

where the sign $*$ denotes convolution operation over time. Our bivariate probability distributions are zero for $t < 0$, so the convolution in (66) is understood as follows

$$\left(\mathbb{P}_{iq}^{(\mu-1)} * \mathbb{P}_{qj}^{(1)}\right)(t) = \int_0^t \mathbb{P}_{iq}^{(\mu-1)}(t')\, \mathbb{P}_{qj}^{(1)}(t-t')\, dt' \tag{67}$$

In these equations, the moment $t = 0$ represents the moment when external factors cause the state *i* of MEE to arise. Equation (66) does not imply any additional restrictions on the type of time dependence of the probability distributions $\mathbb{P}_{ij}^{(\mu)}(t)$ for any value of *μ*. In particular, they can be memoryless distributions, as is required in CTMC, but this is not necessary here.

### ***3.5 Calculation of MCS fluorescence parameters after pulse excitation***

The fluorescence emission observed at time *t* after pulse excitation at time $t = 0$ can be understood as a composite of emissions of excited molecules described by different step number values. The emission of molecules directly excited by an external light source is assigned a step number value of 1, the emission of molecules that have obtained their excitation as a result of a single transfer from directly excited molecules is assigned a step number value of 2, the emission of molecules that obtained their excitation by a two-step transfer from directly excited molecules is assigned a step number value of 3, and so on. It follows from relation (66) that the SPDFs describing the time dependencies of the transition probability densities of MEE to individual chain states in *μ* steps are given by the elements of the $\mathbb{P}^{(\mu)}(t)$ matrix, which is the *μ*th convolution power over time of the $\mathbb{P}^{(1)}(t)$ matrix

$$\mathbb{P}^{(\mu)}(t) = \left(\mathbb{P}^{(1)}\right)^{*\mu}(t) \equiv \underbrace{\left(\mathbb{P}^{(1)} * \mathbb{P}^{(1)} * \ldots * \mathbb{P}^{(1)}\right)}_{\mu}(t) \tag{68}$$

The temporal dependence of the total intensity of visits to state *j* by MEE placed at time $t = 0$ in state *i* is described by the element $\mathbb{P}_{ij}(t)$ of f the matrix $\mathbb{P}(t)$ given by the expression

$$\mathbb{P}(t) = I_\nu \delta(t) + \mathbb{P}^{(1)}(t) + \left(\mathbb{P}^{(1)}\right)^{*2}(t) + \left(\mathbb{P}^{(1)}\right)^{*3}(t) + \dots \tag{69}$$

where $\delta(t)$ is the Dirac delta function. The component $I_\nu \delta(t)$ determines the intensity of MEE visits caused by direct generation of excited states by external light. We will refer to the matrix $\mathbb{P}(t)$ as the matrix of intensity of MEE visits to individual states.

As can be seen from Eq. (69), in our approach there is no strict correspondence between the step number and the real time. According to this equation, at any instant *t* in the considered system there can exist simultaneously an essentially unlimited number of excited molecules characterized by different values of the step number. The essence of the significance of the matrix $\mathbb{P}(t)$ lies in the fact that its elements describe the average properties of all MEEs in the considered MCS at time *t*, taking NET processes into account. Since the direct calculation of the matrix $\mathbb{P}(t)$ as an infinite series of multiple convolution integrals is not possible, it is reasonable to proceed further by considering the matrix $\hat{\mathbb{P}}(s)$ which is the elementwise Laplace transform of the matrix $\mathbb{P}(t)$

$$\hat{\mathbb{P}}(s) = \left[\hat{\mathbb{P}}_{ij}(s)\right]_{\nu\times\nu} = \mathcal{L}\left(\mathbb{P}(t)\right) \tag{70}$$

where

$$\mathcal{L}\left(f(t)\right) = \int_0^\infty \mathrm{e}^{-st} f(t)\, dt \tag{71}$$

Because

$$\mathcal{L}\left[\left(\mathbb{P}^{(1)}\right)^{*\mu}(t)\right] = \left(\hat{\mathbb{P}}^{(1)}\right)^{\mu}(s) \tag{72}$$

it follows from Eq. (69) that

$$\hat{\mathbb{P}}(s) = \sum_{\mu=0}^{\infty}\left(\hat{\mathbb{P}}^{(1)}(s)\right)^{\mu} \tag{73}$$

Thanks to the Laplace transformation, it becomes possible to replace the convolution powers of $\mathbb{P}^{(1)}(t)$ by the ordinary powers of $\hat{\mathbb{P}}^{(1)}(s)$. In this work we deal with systems for which the functions $\Theta_{ij}^{(1)}(t)$, $\Phi_i^{(1)}(t)$, and $Z_i^{(1)}(t)$, which are elements of the matrix $\mathbb{P}^{(1)}(t)$, are given by Eqs. (23)-(29). This means that these functions are integrable in the interval $[0,\infty)$, and this allows us to conclude that $\mathcal{L}$-transforms of these functions, $\hat{\Theta}_{ij}^{(1)}(s)$, $\hat{\Phi}_i^{(1)}(s)$, and $\hat{Z}_i^{(1)}(s)$ exist, and thus the $\hat{\mathbb{P}}(s)$ transform exists. The elements of the matrix $\hat{\mathbb{P}}^{(1)}(s)$ are by definition nonnegative. Furthermore, based on Eqs. (31)-(34) and using the definition of the Laplace transform, one can conclude that these elements are also less than unity. Hence we can expect $\lim_{k\to\infty}\left(\hat{\mathbb{P}}^{(1)}\right)^{k}(s) = 0$, i.e., that the series (73) converges. Under these conditions, Equation (73) can be written as [56,65,66]

$$\hat{\mathbb{P}}(s) = \left(I_\nu - \hat{\mathbb{P}}^{(1)}(s)\right)^{-1} \tag{74}$$

Before calculating the matrix $\hat{\mathbb{P}}(s)$, let us briefly analyze the properties of the matrix $\hat{\mathbb{P}}^{(1)}(s)$. It can see from Eqs. (59) and (70) that structure of matrix $\hat{\mathbb{P}}^{(1)}(s)$ is analogous to the of the matrix $\mathbb{P}^{(1)}(t)$

$$\hat{\mathbb{P}}^{(1)}(s) = \begin{bmatrix} \hat{\Theta}^{(1)}(s) & \hat{\Psi}^{(1)}(s) \\ O_{m\times n} & O_{m\times m} \end{bmatrix} \tag{75}$$

Here, the submatrix $\hat{\Theta}^{(1)}(s)$ describes the time-dependent behavior of the MEE transitions between transient states and is the $\mathcal{L}$-transform of the matrix $\Theta^{(1)}(t)$ given by Eq. (24). It follows from Eqs. (42) and (71) that if the energy transfer is FRET, then the elements of the $\hat{\Theta}^{(1)}(s)$ submatrix can be given by

$$\hat{\Theta}_{ij}^{(1)}(s) = \frac{\gamma_{ij}}{\Gamma_i} f_{\mathrm{F}}\left[x_i(s)\right] \tag{76}$$

where

$$x_i(s) = \frac{\Gamma_i}{\sqrt{s\,\tau_{0i} + 1}} \tag{77}$$

In Eq. (75) the submatrix $\hat{\Psi}^{(1)}(s)$ describes the temporal dependencies of the MEE transitions to the terminating states and is the $\mathcal{L}$-transform of the submatrix $\Psi^{(1)}(t)$ present in Eq. (59). If $\Psi^{(1)}(t)$ is structured according to expression (60), then $\hat{\Psi}^{(1)}(s)$ has the form

$$\hat{\Psi}^{(1)}(s) = \left[\hat{\Phi}^{(1)}(s) \quad \hat{Z}^{(1)}(s)\right] \tag{78}$$

Using (25) (43) and (71) we get

$$\hat{\Phi}^{(1)}(s) = \mathrm{diag}\left(\hat{\Phi}_1^{(1)}(s), \dots, \hat{\Phi}_n^{(1)}(s)\right) \tag{79}$$

where

$$\hat{\Phi}_i^{(1)}(s) = \phi_{0i}\,\frac{1 - f_{\mathrm{F}}\left[x_i(s)\right]}{s\,\tau_{0i} + 1} \tag{80}$$

It follows from Eqs. (27) and (29) that

$$\hat{Z}^{(1)}(s) = \mathrm{diag}\left(\hat{Z}_1^{(1)}(s), \dots, \hat{Z}_n^{(1)}(s)\right) \tag{81}$$

where

$$\hat{Z}_i^{(1)}(s) = \frac{1 - \phi_{0i}}{\phi_{0i}}\,\hat{\Phi}_i^{(1)}(s) \tag{82}$$

It should be noted that expressions (76), (77), and (80) are valid under the assumption that the intermolecular distances of closest approach, $a_{ij}$, are negligibly small, and that the transfer rate $\mathbb{K}_{ij}^{(1)}(t)$ is given by Eq. (38). If we wish to assume $a_{ij} > 0$, then for FRET the rate $\mathbb{K}_{ij}^{(1)}(t)$ is given by expression (36). In this case, analytical expressions for the transforms $\hat{\Theta}_{ij}^{(1)}(s)$, $\hat{\Phi}_i^{(1)}(s)$, and $\hat{Z}_i^{(1)}(s)$ are not available, which does not preclude the possibility of solving this problem numerically. An important property of the matrix $\hat{\mathbb{P}}^{(1)}(s)$ is the values of the transforms $\hat{\Theta}_{ij}^{(1)}(s)$, $\hat{\Phi}_i^{(1)}(s)$, and $\hat{Z}_i^{(1)}(s)$ for $s = 0$. Based on Eqs. (31)-(33), the following relationships occur

$$\hat{\Theta}_{ij}^{(1)}(s=0) = \theta_{ij}^{(1)} \tag{83}$$

$$\hat{\Phi}_i^{(1)}(s=0) = \phi_i^{(1)} \tag{84}$$

$$\hat{Z}_i^{(1)}(s=0) = \zeta_i^{(1)} \tag{85}$$

It can be shown [66] that if $\hat{\mathbb{P}}^{(1)}(s)$ is given by expression (75), then the matrix $\hat{\mathbb{P}}(s)$ has a block structure analogous to that of the matrix $\hat{\mathbb{P}}^{(1)}(s)$

$$\hat{\mathbb{P}}(s)=\begin{bmatrix}\hat{\Theta}(s) & \hat{\Psi}(s)\\ \mathrm{O}_{m\times n} & I_m\end{bmatrix} \tag{86}$$

where the $n\times n$ matrix $\hat{\Theta}(s)$ and the $n\times m$ matrix $\hat{\Psi}(s)$ are defined by the relations

$$\hat{\Theta}(s)=\left(I_n-\hat{\Theta}^{(1)}(s)\right)^{-1} \tag{87}$$

$$\hat{\Psi}(s)=\hat{\Theta}(s)\hat{\Psi}^{(1)}(s) \tag{88}$$

If the $\hat{\Psi}^{(1)}(s)$ submatrix is structured as in expression (78), then

$$\hat{\Psi}(s)=\begin{bmatrix}\hat{\Phi}(s) & \hat{Z}(s)\end{bmatrix} \tag{89}$$

where

$$\hat{\Phi}(s)=\hat{\Theta}(s)\hat{\Phi}^{(1)}(s) \tag{90}$$

$$\hat{Z}(s)=\hat{\Theta}(s)\hat{Z}^{(1)}(s) \tag{91}$$

After finding $\hat{\mathbb{P}}(s)$, using the equations (86)-(91), we can calculate the matrix $\mathbb{P}(t)$ by the inverse Laplace transform $\hat{\mathbb{P}}(s)$, element by element

$$\mathbb{P}(t)=\mathcal{L}^{-1}\left(\hat{\mathbb{P}}(s)\right) \tag{92}$$

From Eq. (86), it follows that the matrix $\mathbb{P}(t)$ is of the form

$$\mathbb{P}(t)=\begin{bmatrix}\Theta(t) & \Psi(t)\\ O_{m\times n} & I_m\delta(t)\end{bmatrix} \tag{93}$$

where $\Theta(t)$ and $\Psi(t)$ can be calculated by the analytical or numerical inverse Laplace transformation of the $\hat{\Theta}(s)$ and $\hat{\Psi}(s)$ submatrices given by Eqs. (87) and (88), respectively. By analogy with the $\hat{\Psi}(s)$ submatrix, the structure of the $\Psi(t)$ submatrix is closely dependent on the structure of the $\hat{\Psi}^{(1)}(s)$ submatrix. If $\hat{\Psi}^{(1)}(s)$ is structured as in Eq. (78), then $\Psi(t)$ contains two subsubmatrices, $\Phi(t)$ and $Z(t)$, according to the scheme

$$\Psi(t)=\begin{bmatrix}\Phi(t) & Z(t)\end{bmatrix} \tag{94}$$

The time courses of the elements of the $\Phi(t)$ and $Z(t)$ submatrices can be calculated by the inverse $\mathcal{L}$-transformation of the $\hat{\Phi}(s)$ and $\hat{Z}(s)$ submatrices given by Eqs. (90) and (91), respectively. Using (82) we can also write

$$Z(t)=\frac{1-\phi_{0i}}{\phi_{0i}}\Phi(t) \tag{95}$$

It follows from Eqs. (90) that the subsubmatrix $\Phi(t)$ can be represented as convolution

$$\Phi(t)=\left(\Theta*\Phi^{(1)}\right)(t) \tag{96}$$

We conclude that the fluorescence of a multicomponent system can be understood as the result of the emission of light quanta by a hypothetical set of molecules among which no excitation energy transfer occurs and which are subject only to one-step processes described by the SPDFs contained in the diagonal matrices $\Phi^{(1)}(t)$ and $\hat{Z}^{(1)}(t)$. In this hypothetical set of molecules, the combined effect of the $\delta(t)$ excitation pulse and the subsequent RNET is replaced by a modified excitation run whose shape varies depending on which component of the system is excited and which is observed. All the necessary excitation runs are described by the elements of $\Theta(t)$ matrix.

In the case where NET is determined by dipole-dipole interaction, the necessary elements of the $\hat{\Theta}^{(1)}(s)$, $\hat{\Phi}^{(1)}(s)$, and $\hat{Z}^{(1)}(s)$ matrices can be found from Eqs. (76), (80), and (82). If the elements of the $\hat{\mathbb{P}}^{(1)}(s)$ matrix are numbers, then the results of expressions (87)-(95) are numbers that are the values of the elements of the $\Theta(t)$, $\Phi(t)$, and $Z(t)$ matrices, while if the elements of the $\hat{\mathbb{P}}^{(1)}(s)$ matrix are expressions, then the results of the calculations are expressions that can, for example, be used to calculate the values of the elements of the aforementioned three matrices.

The $\Theta(t)$, $\Phi(t)$, and $Z(t)$ matrices contain important information about the dynamics of excitation energy wandering in the multicomponent system. A brief description of the physical significance of the elements of these matrices is given in the summary below.

$\Theta(t)$ - The value of the element $\Theta_{ij}(t)$ is equal to the average intensity of MEE transfers (average number of transfers per unit time) to molecules of the *j*th component at time *t*, caused by the excitation of a single random molecule of the *i*th component at time $t=0$. The initial act of excitation of the molecule of the *i*th component is also treated here as a visit.

$\Phi(t)$ - The element $\Phi_{ij}(t)$ of the matrix $\Phi(t)$ has the meaning that the product $\Phi_{ij}(t)dt$ is equal to the probability that a light quantum will be emitted by the *j*th component molecule in the time interval $(t,t+dt)$ assuming that the *i*th component molecule was excited by the excitation light at time $t=0$.

$Z(t)$ - The element $Z_{ij}(t)$ of the matrix $Z(t)$ has the meaning that the product $Z_{ij}(t)dt$ is equal to the probability that the MEE generated by a single quantum of excitation light on the *i*th component molecule will be converted to heat on the *j*th component molecule in the time interval $(t,t+dt)$.

Since in our system each individual excitation ends with the transfer of that excitation to the terminating state, it seems reasonable to say that the time integral taken from zero to infinity from the total sum of elements in each row of the $\Phi(t)+Z(t)$ matrix is equal to unity

$$\int_0^\infty \sum_{j=1}^{n}\left(\Phi_{ij}(t)+Z_{ij}(t)\right)dt=1 \tag{97}$$

This is easily proved for a two-component system by using Eqs. (160), and (30); however, for a larger number of components, the proof becomes tedious.

### 3.6 *Calculation of steady-state MCS fluorescence parameters*

As in the case of impulse excitation, the matrix $P$ describing the average number of MEE visits to individual MEE states in MCS, resulting from continuous excitation, can be understood as the sum of matrices describing the probabilities of transition to these states as a result of all possible numbers of steps. From relation (65), it follows that the probabilities of MEE visiting particular states as a result of exactly $\mu$ steps are described by the elements of the matrix, which is the $\mu$th power of the matrix $P^{(1)}$ defined by equations (55), (63), and (64). Hence, we conclude that the matrix $P$ is given by the expression

$$P=\sum_{\mu=0}^{\infty}\left(P^{(1)}\right)^{\mu} \tag{98}$$

It can be seen from Eq. (98) that the sum $I_{\nu}+P$ constitutes a Neumann series. Since all elements of the $P^{(1)}$ matrix are positive and smaller than unity, we can suppose that $\lim_{k\to\infty}\left(P^{(1)}\right)^{k}=0$ occurs in our case, and that the series (98) converges. Then expression (98) can be written in the form [65]

$$P=\left(I_{\nu}-P^{(1)}\right)^{-1} \tag{99}$$

If the matrix $I_n-\theta^{(1)}$ calculated from the matrix $P^{(1)}$ given by expression (63) is non-singular, then it can be shown [66] that the matrix $P$ has a block structure

$$P=\begin{bmatrix}\theta & \psi\\ O_{m\times n} & I_m\end{bmatrix} \tag{100}$$

where

$$\theta=\left(I_n-\theta^{(1)}\right)^{-1} \tag{101}$$

$$\psi=\theta\psi^{(1)} \tag{102}$$

If the $\psi^{(1)}$ matrix has the structure described by equation (64), then the $\psi$ matrix takes the form

$$\psi=\begin{bmatrix}\phi & \zeta\end{bmatrix} \tag{103}$$

From Eqs. (102), (103) and (64) it follows that

$$\phi=\theta\phi^{(1)} \tag{104}$$

$$\zeta=\theta\zeta^{(1)} \tag{105}$$

On the other hand, it follows from the dependencies (54) and (71) that

$$P^{(1)}=\int_0^{\infty}\mathbb{P}^{(1)}(t)\,dt=\hat{\mathbb{P}}^{(1)}(s=0) \tag{106}$$

After inserting this result into Eq. (98) and using (73), we get

$$P=\int_0^{\infty}\mathbb{P}(t)\,dt=\hat{\mathbb{P}}(s=0) \tag{107}$$

The above equation allows us to determine the values of the elements of the $P$ matrix when the $\mathbb{P}(t)$ or $\hat{\mathbb{P}}(s)$ matrix is known. If the structure of the matrix $\mathbb{P}(t)$ is blocky and given by Eqs. (93) and (94), it follows from Eq. (107) that the matrix $P$ also has a blocky structure given by Eq. (100). Thus, the following relationships take place

$$\theta=\int_0^{\infty}\Theta(t)\,dt=\hat{\Theta}(s=0) \tag{108}$$

$$\psi=\int_0^{\infty}\Psi(t)\,dt=\hat{\Psi}(s=0) \tag{109}$$

$$\phi=\int_0^{\infty}\Phi(t)\,dt=\hat{\Phi}(s=0) \tag{110}$$

$$\zeta=\int_0^{\infty}Z(t)\,dt=\hat{Z}(s=0) \tag{111}$$

Expressions (108), (110), and (111) can be useful for finding the elements of the $\theta$, $\phi$, and $\zeta$ submatrices, especially when we know the expressions describing the elements of the $\hat{\Theta}(s)$, $\hat{\Phi}(s)$, and/or $\hat{Z}(s)$ submatrices. The $\theta$, $\phi$, and $\zeta$ submatrices are less significant than the $\Theta(t)$, $\Phi(t)$, and/or $Z(t)$ submatrices, but they still contain important information about the excitation energy wandering in the multicomponent system under consideration. Their physical meanings are discussed in the summary below.

- $\theta$ - Based on [67-69], it can be concluded that the $\theta_{ij}$ element of the $\theta$ submatrix represents the average number of MEE visits to the $j$th component molecules before its final transfer to the terminating state, assuming that the originally excited molecule belonged to the $i$th component. The act of primary excitation by light of the $i$th component molecule is included in the number of visits. Similarly, the sum of the elements contained in the $i$th row of the $\theta$ submatrix determines the average number of MEE visits to all components of the system before it reaches one of the terminating states, assuming that light excitation caused the formation of MEE in state $i$.
- $\phi$ - The $\phi_{ij}$ element of the $\phi$ submatrix is the probability that the MEE delivered to the $i$th component molecule will be converted to a photon on the $j$th component molecule, taking into account the processes of spontaneous emission, internal conversion, and RNET. The $\phi$ submatrix has the same meaning as that defined by Eq. (2).
- $\zeta$ - The $\zeta_{ij}$ element of the $\zeta$ submatrix is the probability that the MEE delivered to the $i$th component molecule will be converted to heat on the $j$th component molecule, taking into account the processes of spontaneous emission, internal conversion, and RNET.

In order to use Eqs. (101), (104), and (105) to calculate the $\theta$, $\phi$, and $\zeta$ submatrices, the elements of the $P^{(1)}$ matrix must be known. In the case where NET is determined by dipole-dipole interaction, the necessary elements of the $\theta^{(1)}$, $\phi^{(1)}$, and $\zeta^{(1)}$ submatrices can be found from Eqs. (45), (47), and (48). If the elements of matrix $P^{(1)}$ are numbers, then the result of expressions (101), (104). and (105) are numbers that are the values of the elements of matrices $\theta$, $\phi$, and $\zeta$. On the other hand, if the elements of the $P^{(1)}$ matrix are expressions, then the results of the calculations are expressions that can, for example, be used to calculate the values of the elements of $\theta$, $\phi$, and $\zeta$ submatrices. It is worth noting that with respect to the elements of the $\phi$ and $\zeta$ submatrices, Eq. (97) takes the form

$$\sum_{j=1}^{n}\left(\phi_{ij}+\zeta_{ij}\right)=1 \tag{112}$$

Intuitively, this equation follows from the assumption that every MCS excitation, sooner or later, ends with the transfer of excitation energy to the terminating state. Strict justification of Eq. (112), however, seems difficult.

### 3.7 Finding the residence probability matrix $\mathcal{P}(t)$

In the considerations carried out within the framework of CTMC, of fundamental importance is the so-called transition function [44], Markov state density function [70], or transition probability function [64,71]. This function is represented by a square matrix $\mathcal{P}(t)$, in which a given element $\mathcal{P}_{ij}(t)$ represents the probability of the system being in state *j* at time *t*, assuming that at time $t=0$ the system was in state *i*. In our case of MCS fluorescence, the $\mathcal{P}(t)$ matrix is $(\nu\times\nu)$-dimensional, and its element $\mathcal{P}_{ij}(t)$ determines the probability of MEE residence at time *t* on the molecule of component *j* if the initial excitation was placed on the molecule of component *i* at time $t=0$

$$\mathcal{P}(t)=\left[\mathcal{P}_{ij}(t)\right]_{\nu\times\nu} \tag{113}$$

We will begin calculating the value of the elements of the matrix $\mathcal{P}(t)$ by introducing the $\nu$-element diagonal matrix $\mathcal{P}^{(1)}(t)$ with the structure

$$\mathcal{P}^{(1)}(t)=\operatorname{diag}\left(\mathcal{P}_{1}^{(1)}(t),\ldots,\mathcal{P}_{\nu}^{(1)}(t)\right) \tag{114}$$

We will assume that the element $\mathcal{P}_{i}^{(1)}(t)$ of this matrix has the meaning of the one-step probability of survival of the object of the chain in the *i*th state until time *t*, assuming that this object began its stay in this state at time $t=0$. It is easy to see that in the case of MCS, the first *n* elements of matrix (114) are the same as the elements of matrix $\mathcal{P}_{\mathrm{M}}^{(1)}(t)$ defined by expression (15). Taking into account the properties of the matrices $\mathcal{P}^{(1)}(t)$ and $\mathbb{P}(t)$, we conclude that the elements $\mathcal{P}_{ij}(t)$ of the matrix (113) can be calculated as

$$\mathcal{P}_{ij}(t)=\left(\mathbb{P}_{ij}*\mathcal{P}_{j}^{(1)}\right)(t) \tag{115}$$

Given that the matrix $\mathcal{P}^{(1)}(t)$ is diagonal, the entire $\mathcal{P}(t)$ matrix can be represented as

$$\mathcal{P}(t)=\left(\mathbb{P}*\mathcal{P}^{(1)}\right)(t) \tag{116}$$

When using the $\mathcal{L}$-transforms $\hat{\mathcal{P}}(s)$ and $\hat{\mathcal{P}}^{(1)}(s)$, the convolution of the matrix can be replaced by the simple product of the transforms

$$\hat{\mathcal{P}}(s)=\hat{\mathbb{P}}(s)\hat{\mathcal{P}}^{(1)}(s) \tag{117}$$

#### 3.7.1 $\mathcal{P}(t)$ matrix for fluorescent MCS

In the case of MCS containing *n* types of fluorescent molecules, which are assigned *n* transient states and *m* terminating states, the first *n* elements of the matrix (114) are the same as the elements of the $\mathcal{P}_{\mathrm{M}}^{(1)}(t)$ matrix given by Eq. (15), and the remaining *m* elements are units for all times. The latter occurs because once the MEE is brought to a terminating state at time $t=0$, it cannot transition to other states and therefore remains in this state forever. Then, the matrix $\mathcal{P}^{(1)}(t)$ can be represented as

$$\mathcal{P}^{(1)}(t)=\begin{bmatrix}\left[\mathcal{P}_{\mathrm{M}}^{(1)}(t)\right]_{n\times n} & \mathrm{O}_{n\times m}\\ \mathrm{O}_{m\times n} & I_m\,H(t)\end{bmatrix} \tag{118}$$

where *H*(*t*) is the Heaviside function. Using Eqs. (117), (86), (87) and (118), we obtain

$$\hat{\mathcal{P}}(s)=\begin{bmatrix}\hat{\Theta}(s)\hat{\mathcal{P}}_{\mathrm{M}}^{(1)}(s) & s^{-1}\hat{\Psi}(s)\\ \mathrm{O}_{m\times n} & s^{-1}I_m\end{bmatrix} \tag{119}$$

where $\hat{\Theta}(s)$ is given by Eq. (87), and $\hat{\Psi}(s)$ is given by Eqs. (89)-(91). From Eq. (119), we conclude that the matrix $\mathcal{P}(t)$ has a block structure given by the diagram

$$\mathcal{P}(t)=\begin{bmatrix}\left[\mathcal{P}_{\mathrm{M}}(t)\right]_{n\times n} & \left[\mathcal{P}_{\Psi}(t)\right]_{n\times m}\\ \mathrm{O}_{m\times n} & I_m\,H(t)\end{bmatrix} \tag{120}$$

In (120), the $\mathcal{P}_{\mathrm{M}}(t)$ submatrix is given by the expression

$$\mathcal{P}_{\mathrm{M}}(t)=\left(\hat{\Theta}*\mathcal{P}_{\mathrm{M}}^{(1)}\right)(t) \tag{121}$$

and the $\mathcal{P}_{\Psi}(t)$ submatrix is given by

$$\mathcal{P}_{\Psi}(t)=\int_{0}^{t}\Psi(u)du \tag{122}$$

The element $\mathcal{P}_{\mathrm{M}ij}(t)$, $i,j\in\{1,\ldots,n\}$ of the submatrix $\mathcal{P}_{\mathrm{M}}(t)$ is the probability of finding MEE at time *t* on molecule of component *j*, if at time $t=0$ this MEE was placed on molecule of component *i*. Similarly, the element $\mathcal{P}_{\Psi ij}(t)$ of the submatrix $\mathcal{P}_{\Psi}(t)$ is the probability of converted MEE being in the *j*th, $j\in\{1,\ldots,m\}$ terminating state at time *t*, if at time $t=0$ this MEE was placed on the molecule of component *i*, $i\in\{1,\ldots,n\}$.

Functions $\mathcal{P}_{\mathrm{M}}(t)$ are fundamental to the CTMC method. However, it must be admitted that this method significantly limits the acceptable range of $\mathcal{P}_{\mathrm{M}}(t)$ functional forms for any Markov process [70]. This limitation is due to the fact that only those $\mathcal{P}_{\mathrm{M}}(t)$ functions are accepted in CTMC that fulfill the Chapman-Kolmogorov equation [44,64,72-76]

$$\mathcal{P}_{\mathrm{M}}(t+u)=\mathcal{P}_{\mathrm{M}}(t)\mathcal{P}_{\mathrm{M}}(u) \tag{123}$$

Within the framework of TRMC, equation (123) may or may not be fulfilled.

### 3.8 Discrete time vs. continuous time in the Markov chain formalism

The results of this work indicate that using the term "time" to describe the step number in DTMC is not justified. The step number cannot be time because it is a dimensionless quantity and has no time dimension. The practice of calling a step number a discrete time leads to the misconception that DTMC contains time, when it does not. As we emphasized at the end of Section 3.3, time is unspecified here. The difference between step number and time is clearly visible in the TRMC model introduced in this work, where the two quantities appear independently of each other. It can be surmised that the argument for the identification of step number and time became the Chapman-Kolmogorov equations. In the case of DTMC, these equations follow from relation (65) and take the form [46,71-73,75,77]

$$P_{ij}^{(\mu_1+\mu_2)}=\sum_{q=1}^{\nu}P_{iq}^{(\mu_1)}\,P_{qj}^{(\mu_2)} \tag{124}$$

or in matrix form

$$P^{(\mu_1+\mu_2)}=P^{(\mu_1)}P^{(\mu_2)} \tag{125}$$

In the case of CTMC, on the other hand, expression (123) was adopted as an important form of the Chapman-Kolmogorov equation. Despite the great similarity of the Chapman-Kolmogorov equations in DTMC and CTMC, the probability matrices $P^{(\mu)}$ and $\mathcal{P}(t)$ considered by these equations have different physical meanings. Therefore, this similarity cannot provide a basis for conclusions about the existence of a relationship between step number and real time. Postulating the fulfillment of equation (123) within CTMC framework has only partially solved the problem of time in Markov processes. As already mentioned, the adoption of this postulate has meant that within the CTMC framework, only processes in which the temporal probability density distributions of transitions between states are exclusively continuous and exclusively exponential can be considered.

Within the TRMC model introduced in this work, real time appears next to "discrete time" (actually: step number), rather than as an extension of "discrete time" to the $\mathbb{R}_+^0$ region. In this new approach, the Chapman-Kolmogorov equations in the version of expressions (123) and (125) lose their fundamental meaning. This makes it unnecessary to use transition rate matrix $Q$ under TRMC framework, unlike CTMC, and the temporal event distribution functions do not have to be exponential and can be arbitrary.

The TRMC model we introduced, in addition to using continuous time as, for example, in expressions (42)-(44), can also use discrete time, for example, by assuming that for $t\geq 0$, the elements of the $\mathbb{P}^{(1)}(t)$ matrix are of the form

$$\mathbb{P}_{ij}^{(1)}(t)=\sum_{k=1}^{p}\;A_{ijk}^{(1)}\,\delta(t-t_{ijk}^{(1)}) \tag{126}$$

where $A_{ijk}^{(1)}$ are the probabilities of a one-step transition of the described system from state $i$ to state $j$ at time $t_{ijk}^{(1)}$, $p$ is an integer greater than zero, and $\delta(t)$ is the Dirac delta function. We assume that the values of probabilities $A_{ijk}^{(1)}$ satisfy Eq. (61). Note, however, that in Eq. (126) it is actually not the time that is discrete, but the temporal probability density distributions of transitions between states.

### *3.9 The relationship of TRMC and TUMC to renewal theory*

The issue of theoretical description of RNET in MCS discussed in this paper can also be analyzed using expressions derived from renewal theory (RT) [47-51]. Different versions of this theory allow us to describe the renewal processes, semi-Markov renewal processes, as well as to refer to earlier described processes such as DTMC and CTMC. In terms of homogeneous renewal process, the non-decreasing real matrix function $Q(t)$, known as the renewal kernel, is of fundamental importance. The individual elements $Q_{ij}(t)$ of this matrix represent the subnormalized cumulative distribution functions (SCDFs) of a single-step transition of the system from state $i$ to state $j$. It is possible to determine the relationship between the renewal kernel and the $\mathbb{P}^{(1)}(t)$ matrix defined in this work by Eq. (59). If we assume that the matrices $Q(t)$ and $\mathbb{P}^{(1)}(t)$ describe the same stochastic process, then we can write

$$Q(t)=\int_0^t\mathbb{P}^{(1)}(u)du \tag{127}$$

To understand the need to use the term "subnormalized" when describing the properties of elements of the matrix $Q(t)$, note that for $t\to\infty$, the functions $Q_{ij}(t)$ generally take values that are less than unity. This follows, for example, from Eqs. (127) and (106), which imply

$$\lim_{t\to\infty}Q(t)=P^{(1)} \tag{128}$$

and from Eq. (61), which implies $P_{ij}^{(1)}\leq 1$.

Among the results of RT considerations, one of the main objects of interest is the matrix $R(t)$, called the renewal function of the process [48] or the Markov renewal matrix [49-51]. The element $R_{ij}(t)$ of the renewal matrix determines the number of visits to the state $j$ up to time $t$, provided that at time $t=0$ the system was in state $i$, whereby the matrix $R(t)$ is expressed as the sum of all convolutional powers of the matrix $Q(t)$

$$R(t)=\sum_{\mu=0}^{\infty}Q^{\bullet\mu}(t) \tag{129}$$

Equation (129) takes into account not only the transfer from state $i$ to state $j$, but also the fact of external creation of state $j=i$ at time $t=0$. The latter is described by the first term of the above sum, specified by the value $\mu=0$, and which is defined by the expression

$$Q^{\bullet 0}(t)=I_\nu\,H(t) \tag{130}$$

where $H(t)$ is the Heaviside step function. For $\mu>0$, the terms $Q^{\bullet\mu}(t)$ are calculated from the expression [50,51]

$$Q^{\bullet\mu}(t)=\int_0^t Q^{\bullet(\mu-1)}(t-u)\,dQ(u) \tag{131}$$

From Eqs. (131) and (127), it follows that

$$\lim_{t\to\infty}Q^{\bullet\mu}(t)=\begin{cases}I_\nu & \text{if } \mu=0\\ \left(P^{(1)}\right)^{\mu} & \text{if } \mu>0\end{cases} \tag{132}$$

which, when substituted into (129) and using (98), gives

$$\lim_{t\to\infty}R(t)=\sum_{\mu=1}^{\infty}\left(P^{(1)}\right)^{\mu}=P \tag{133}$$

Equations (128) and (133) show that, at the $t\to\infty$ boundary,

the description of a given stochastic process using RT corresponds to the description of this process using TUMC/DTMC. With regard to the function $R(t)$, it is important [48,49,51] that after applying the Laplace-Stieltjes (LS) transform of the form

$$\bar{Q}(s) = \int_0^\infty e^{-st} dQ(t) \tag{134}$$

within RT, one obtains

$$\bar{R}(s) = \sum_{\mu=0}^{\infty} \left[ \bar{Q}(s) \right]^{\mu} \tag{135}$$

Considering that the right-hand side of Eq. (135) constitutes a Neumann series, this equation can be written in the form

$$\bar{R}(s) = \left( I - \bar{Q}(s) \right)^{-1} \tag{136}$$

Despite such a possibility, it is rather rare in RT to find values of the function $R(t)$ by inverse LS transformation of the function $\bar{R}(s)$ to time space. This is probably because the technique of numerical calculation of inverse LS transforms is poorly developed compared to, for example, the technique of numerical calculation of inverse $\mathcal{L}$ transforms. Therefore, in RT, the function $R(t)$ is more often found rather by solving an integral equation of the form

$$R(t) = I_v H(t) + \int_0^t R(t-u)\, dQ(u) \tag{137}$$

This equation can be obtained by the inverse LS transformation of the equation

$$\bar{R}(s) = I_v + \bar{Q}(s)\bar{R}(s) \tag{138}$$

resulting from relation (136).

Equations analogous to (138) and (137), but concerning the functions $\hat{\mathbb{P}}(s)$ and $\mathbb{P}(t)$, can also be obtained within the TRMC model proposed in this work. After multiplying both sides of Eq. (74) by $(I_v - \hat{\mathbb{P}}^{(1)}(s))$, we obtain the relationship

$$\hat{\mathbb{P}}(s) = I_v + \hat{\mathbb{P}}(s)\hat{\mathbb{P}}^{(1)}(s) \tag{139}$$

which in time space corresponds to the integral equation

$$\mathbb{P}(t) = I_v \delta(t) + \int_0^t \mathbb{P}(t-u)\mathbb{P}^{(1)}(u)du \tag{140}$$

The equation (140) applies within the TRMC model and is equivalent to Eq. (137) obtained within RT. In our opinion, the significance of equation (140) for finding the values of the elements of the $\mathbb{P}(t)$ matrix is not high. For this purpose, it seems much easier to use the relations (74) and (92).

If we consider that for any matrix $A(t)$ with the properties of matrix $Q(t)$ or $R(t)$, the relationship between the LS-transform $\bar{A}(s)$ and the $\mathcal{L}$-transform $\hat{A}(s)$ of the matrix $A(t)$ is given by [78]

$$\bar{A}(s) = s\,\hat{A}(s) \tag{141}$$

then Eq. (127) can be written

$$\bar{Q}(s) = \hat{\mathbb{P}}^{(1)}(s) \tag{142}$$

After inserting (142) into (136) and taking (73) into account, we obtain

$$\hat{R}(s) = s^{-1} \sum_{\mu=0}^{\infty} \left( \hat{\mathbb{P}}^{(1)}(s) \right)^{\mu} = s^{-1}\hat{\mathbb{P}}(s) \tag{143}$$

or

$$R(t) = \int_0^t \mathbb{P}(u)du \tag{144}$$

Equations (127) and (144) show what the direct relationships are between the basic functional matrices characterizing the stochastic process under consideration within RT and the corresponding matrices used within the TRMC model. It is also worth noting that RT equation (136), after using relations (141)-(143) and (127), can be written in the form of our Eq. (74). This leads us to conclude that the results obtained using the TRMC method are equivalent to those obtained using the equations predicted by RT. The integral transformation used in the RT technique is the rarely used LS transformation defined by Eq. (134), and the literature describing the properties of this transformation and the technique for finding inverse LS-transforms is relatively scarce. On the other hand, the integral transformation used in the TRMC technique is the frequently used $\mathcal{L}$-transformation defined by Eq. (71), and the literature describing the properties of this transformation and the technique for finding inverse $\mathcal{L}$-transforms is very rich.

### 3.10 Comparison of selected TRMC results with analogous results obtained by usual analytical methods

In paper [32], it is shown that the fluorescence decay of individual components in an $n$-component system with a random distribution of molecules can be described by a system of equations of the form

$$\hat{N}_i = \hat{\phi}_i^{(0)} \left[ 1 - \hat{A}_{ii} \right]^{-1} \left[ N_{0i} + \sum_{\substack{k=1 \\ k \neq i}}^{n} \left( \hat{\phi}_k^{(0)} \right)^{-1} \hat{A}_{ki}\, \hat{N}_k \right], \quad i = 1, 2, .., n \tag{145}$$

where $N_{0i}$ is the number of excited molecules of component $i$ at time $t = 0$, and $N_i$ is the number of these molecules at any time $t > 0$. In order to be able to compare this result with the results of the present work, it is convenient to introduce the SPDF vector of effective fluorescence, $E(t)$ of the form

$$E(t) = \left[ \sum_{j=1}^{n} X_j\, \Phi_{ji}(t) \right]_{1\times n} = X\,\Phi(t) \tag{146}$$

where $X_i = N_{0i}/N_0$, $N_0 = \sum_{i=1}^{n} N_{0i}$, and $X = \left[ X_i \right]_{1\times n}$. The $E_i(t)$ elements of this vector are such that the product $E_i(t)dt$ is equal to the probability that a photon absorbed by the multicomponent system from an excitation light beam at time $t = 0$, will result in the emission of a quantum of light by a component $i$ molecule in the time interval $(t, t+dt)$. Using the terminology of the present work, such symbols in (145) as $\hat{N}_i$, $\hat{\phi}_i^{(0)}$, $\hat{A}_{ki}$, should be interpreted as follows

$$\hat{N}_i = N_0 \frac{\tau_{0i}}{\phi_{0i}} \hat{E}_i(s) \tag{147}$$

$$\hat{\phi}_i^{(0)} = \frac{\tau_{0i}}{\phi_{0i}} \hat{\Phi}_i^{(1)}(s) \tag{148}$$

$$\hat{A}_{ki} = \hat{\Theta}_{ki}^{(1)}(s) \tag{149}$$

Thus Eq. (145) can be written as

$$\hat{E}_i(s) = \hat{\Phi}_i^{(1)}(s)\left(1-\hat{\Theta}_{ii}^{(1)}(s)\right)^{-1} \times \left( X_i + \sum_{\substack{k=1 \\ k\neq i}}^{n} \left(\hat{\Phi}_k^{(1)}(s)\right)^{-1} \hat{\Theta}_{ki}^{(1)}(s)\, \hat{E}_k(s) \right) \tag{150}$$

After introducing the auxiliary vector $\hat{W}(s)$, such that

$$\hat{E}(s) = \hat{W}(s)\hat{\Phi}^{(1)}(s) \tag{151}$$

where the diagonal matrix $\hat{\Phi}^{(1)}(s)$ is given by Eq. (79), we obtain

$$\hat{W}_i(s)\left(1-\hat{\Theta}_{ii}^{(1)}(s)\right) - \sum_{\substack{k=1 \\ k\neq i}}^{n} \hat{W}_k(s)\hat{\Theta}_{ki}^{(1)}(s) = X_i, \quad i = 1,2,..,n \tag{152}$$

or

$$\hat{W}(s)\left(I_n - \hat{\Theta}^{(1)}(s)\right) = X \tag{153}$$

where $\Theta^{(1)}(s)$ is a Laplace transform of the matrix $\Theta^{(1)}(t)$ is defined by Eq. (23). Using Eqs. (146) and (151), it is easy to convert (153) to the form

$$\hat{\Phi}(s) = \left(I_n - \hat{\Theta}^{(1)}(s)\right)^{-1} \hat{\Phi}^{(1)}(s) \tag{154}$$

which is our Eq. (90). This shows that the results of our TRMC method are consistent with the results obtained for the same issue using conventional analytical methods.

# 4 Examples of applications of TUMC and TRMC methods to describe RNET in simple multicomponent systems

In this section, we will provide examples of using the TRMC and TUMC methods to find expressions describing the elements of the matrices $\hat{\Phi}(s)$ and $\phi$. In Section 4.3, we will give an example of the application of these methods to the analysis of the time properties of absorbing and partially absorbing states.

## 4.1 Two-component fluorescent mixture

Excitation energy transfer in a binary system has already been partially discussed in Section 3.4. A prerequisite for detailed calculations is knowledge of the concentrations of the components $c_1$ and $c_2$, their lifetimes $\tau_{01}$ and $\tau_{02}$, the quantum yields $\phi_{01}$ and $\phi_{02}$, and the four Förster radii $R_{0ij}$, $i,j \in \{1,2\}$. With these data we are able to calculate the values of the Laplace transforms $\hat{\Theta}_{ij}^{(1)}(s)$, $\hat{\Phi}_i^{(1)}(s)$, and $\hat{Z}_i^{(1)}(s)$, ($i,j \in \{1,2\}$) of the SPDFs defined by Eqs. (23)-(28). In the special case where RNET is due to dipole-dipole interaction, these functions are given by Eqs (76), (80), and (82). If our goal is to find functions describing the elements of the matrix $\hat{\Phi}(s)$ then we first create the matrix $\hat{\mathbb{P}}^{(1)}(s)$ defined by Eqs. (75) and (78). The elements of this matrix describe the kinetics of the transitions between the six states of the system shown in Fig. 2.

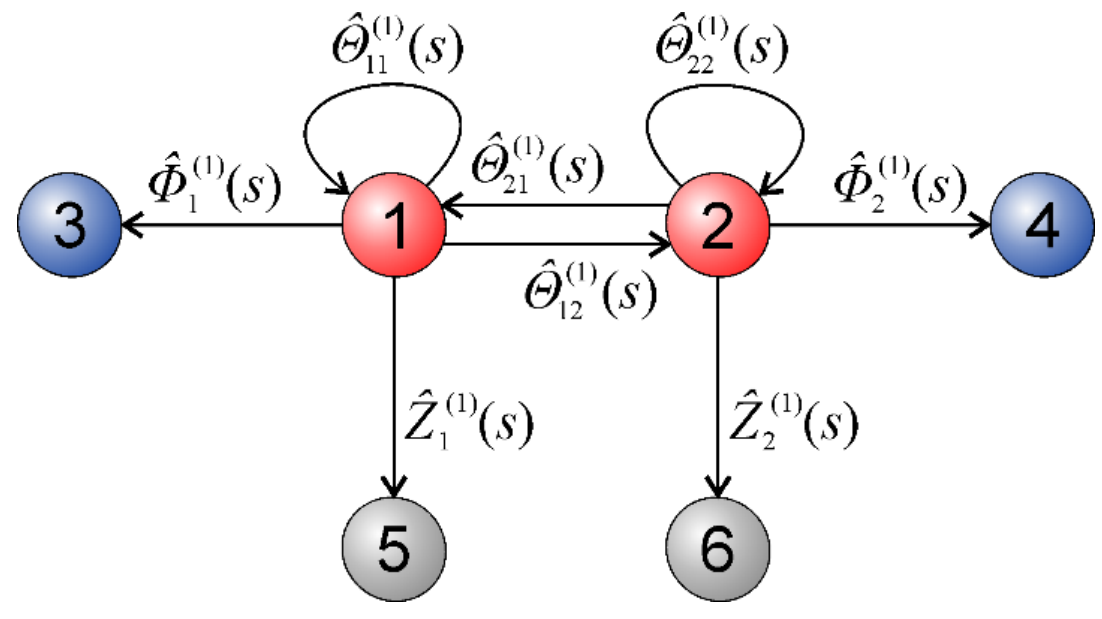

**Figure 2.** Six-state diagram of the Markov process for a binary system. The basic features of this diagram are the same as in Fig. 1. Instead of the $\Theta_{ij}^{(1)}(t)$, $\Phi_i^{(1)}(t)$, and $Z_i^{(1)}(t)$ functions, the Laplace transforms of these functions $\hat{\Theta}_{ij}^{(1)}(s)$, $\hat{\Phi}_i^{(1)}(s)$, and $\hat{Z}_i^{(1)}(s)$ are placed.

In the considered case we have $n = 2$, and the submatrices $\hat{\Theta}^{(1)}(s)$, $\hat{\Phi}^{(1)}(s)$ and $\hat{Z}^{(1)}(s)$ defined by Eqs. (75) and (78) take the forms

$$\hat{\Theta}^{(1)}(s) = \left[\hat{\Theta}_{ij}^{(1)}(s)\right]_{2\times 2} \tag{155}$$

$$\hat{\Phi}^{(1)}(s) = \operatorname{diag}\left(\hat{\Phi}_1^{(1)}, \hat{\Phi}_2^{(1)}\right) \tag{156}$$

$$\hat{Z}^{(1)}(s) = \operatorname{diag}\left(\hat{Z}_1^{(1)}, \hat{Z}_2^{(1)}\right) \tag{157}$$

while the matrix $\hat{\Theta}(s)$ defined by (87) is given by the expression

$$\hat{\Theta}(s) = \left(I_2 - \hat{\Theta}^{(1)}(s)\right)^{-1} = \frac{1}{D}\begin{bmatrix} 1-\hat{\Theta}_{22}^{(1)} & \hat{\Theta}_{12}^{(1)} \\ \hat{\Theta}_{21}^{(1)} & 1-\hat{\Theta}_{11}^{(1)} \end{bmatrix} \tag{158}$$

where

$$D = \left(1-\hat{\Theta}_{11}^{(1)}\right)\left(1-\hat{\Theta}_{22}^{(1)}\right) - \hat{\Theta}_{12}^{(1)}\,\hat{\Theta}_{21}^{(1)} \tag{159}$$

Knowing $\hat{\Theta}(s)$, we can use Eq. (90) to calculate the desired matrix $\hat{\Phi}(s)$

$$\hat{\Phi}(s) = \frac{1}{D}\begin{bmatrix} (1-\hat{\Theta}_{22}^{(1)})\hat{\Phi}_1^{(1)} & \hat{\Theta}_{12}^{(1)}\,\hat{\Phi}_2^{(1)} \\ \hat{\Theta}_{21}^{(1)}\,\hat{\Phi}_1^{(1)} & (1-\hat{\Theta}_{11}^{(1)})\hat{\Phi}_2^{(1)} \end{bmatrix} \tag{160}$$

From the above expression, it can be seen that the course of fluorescence intensity of a binary solution is in general non-exponential, regardless of which component is excited by an external pulse of light. If there is no NET from component 2 to component 1, that is, if $\hat{\Theta}_{21}^{(1)}(s) = 0$, then expression (160) reduces to

$$\hat{\Phi}(s) = \begin{bmatrix} \dfrac{\hat{\Phi}_1^{(1)}}{1-\hat{\Theta}_{11}^{(1)}} & \dfrac{\hat{\Theta}_{12}^{(1)}}{1-\hat{\Theta}_{11}^{(1)}}\dfrac{\hat{\Phi}_2^{(1)}}{1-\hat{\Theta}_{22}^{(1)}} \\ 0 & \dfrac{\hat{\Phi}_2^{(1)}}{1-\hat{\Theta}_{22}^{(1)}} \end{bmatrix} \tag{161}$$

Considering the results given in Appendix A, from Eq. (161) we see that then if only component 2 is excited, then its fluorescence decays exponentially, regardless of the general non-exponentiality of the functions $\Phi^{(1)}_2(t)$ and $\Theta^{(1)}_{22}(t)$. If only component 1 is excited under the same conditions, then one observes fluorescence of both components, whose time courses are generally nonexponential.

For finding the $\phi$ matrix, we can use two methods. The first method concerns the situation when we know the matrix (160). Then it is enough to use expressions (110), (83) and (84). This gives

$$\phi = \frac{1}{d}\begin{bmatrix} (1-\theta^{(1)}_{22})\phi^{(1)}_1 & \theta^{(1)}_{12}\,\phi^{(1)}_2 \\ \theta^{(1)}_{21}\,\phi^{(1)}_1 & (1-\theta^{(1)}_{11})\phi^{(1)}_2 \end{bmatrix} \tag{162}$$

where

$$d = \left(1-\theta^{(1)}_{11}\right)\left(1-\theta^{(1)}_{22}\right) - \theta^{(1)}_{12}\,\theta^{(1)}_{21} \tag{163}$$

The quantities $\theta^{(1)}_{ij}$ and $\phi^{(1)}_i$ occurring here can be calculated, for example, from expressions (45)-(47). The second way to find the eta matrix is to use the TUMC method. We begin by constructing a $P^{(1)}$ matrix such as shown in expression (62) by determining the values of the elements of the $\theta^{(1)}$ , $\phi^{(1)}$, and $\zeta^{(1)}$ submatrices based on Eqs. (63), (64), and (45)-(48). Based on Eq. (101), the matrix $\theta$ is given by

$$\theta = \frac{1}{d}\begin{bmatrix} 1-\theta^{(1)}_{22} & \theta^{(1)}_{12} \\ \theta^{(1)}_{21} & 1-\theta^{(1)}_{11} \end{bmatrix} \tag{164}$$

where $d$ is given by Eq. (163). Knowing the $\theta$ matrix, we can use Eq. (104) to calculate the desired $\phi$ matrix. The result thus obtained is, for obvious reasons, the same as that presented by expression (162).

### 4.2 Three-component fluorescent mixture

In this section, we sketch the course of calculating the elements of the $\hat{\Phi}(s)$ and $\phi$ matrices for the ternary solution when NET is described by expression (38). For the sake of shortening our considerations, we will limit ourselves to using the TRMC method only. A schematic of the NET, spontaneous emission, and internal conversion processes occurring in this system is shown in Fig. 3. The way the states are numbered and the associated colors are the same here as in the previous figures. The quantities $\Theta^{(1)}_{ij}$ are SPDFs describing the kinetics of one-step excitation energy transfer between states 1, 2, and 3. These states correspond to the presence of excitation energy on the molecules of the individual system components. The $\Phi^{(1)}_i$ and $Z^{(1)}_i$ quantities are SPDFs describing the kinetics of one-step spontaneous emission and internal conversion, respectively. We assume that for this system we know the three concentrations $c_i$ of the components, the three lifetimes $\tau_{0i}$, the three quantum yields $\phi_{0i}$, and the nine Förster radii $R_{0ij}$, where $i, j \in \{1,2,3\}$. These data and the expressions (76)-(82) allow us to calculate all the elements of the matrix $\hat{\mathbb{P}}^{(1)}(s)$. Using Eq. (87) we find the elements of the matrix $\hat{\Theta}(s)$. This gives

$$\left.\begin{aligned} \hat{\Theta}_{ii}(s) &= \frac{1}{D}\left[\left(1-\hat{\Theta}^{(1)}_{jj}\right)\left(1-\hat{\Theta}^{(1)}_{kk}\right) - \hat{\Theta}^{(1)}_{jk}\hat{\Theta}^{(1)}_{kj}\right] \\ \hat{\Theta}_{ij}(s) &= \frac{1}{D}\left[\hat{\Theta}^{(1)}_{ij}\left(1-\hat{\Theta}^{(1)}_{kk}\right) + \hat{\Theta}^{(1)}_{ik}\hat{\Theta}^{(1)}_{kj}\right] \end{aligned}\right\} \quad i, j, k = \{1,2,3\}, \quad k \neq j \neq i \tag{165}$$

where

$$\begin{aligned} D = &\left(1-\hat{\Theta}^{(1)}_{11}\right)\left(1-\hat{\Theta}^{(1)}_{22}\right)\left(1-\hat{\Theta}^{(1)}_{33}\right) \\ &-\left(1-\hat{\Theta}^{(1)}_{11}\right)\hat{\Theta}^{(1)}_{12}\hat{\Theta}^{(1)}_{23} - \left(1-\hat{\Theta}^{(1)}_{22}\right)\hat{\Theta}^{(1)}_{13}\hat{\Theta}^{(1)}_{31} \\ &-\left(1-\hat{\Theta}^{(1)}_{33}\right)\hat{\Theta}^{(1)}_{12}\hat{\Theta}^{(1)}_{21} - \hat{\Theta}^{(1)}_{12}\hat{\Theta}^{(1)}_{23}\hat{\Theta}^{(1)}_{31} - \hat{\Theta}^{(1)}_{13}\hat{\Theta}^{(1)}_{32}\hat{\Theta}^{(1)}_{21} \end{aligned} \tag{166}$$

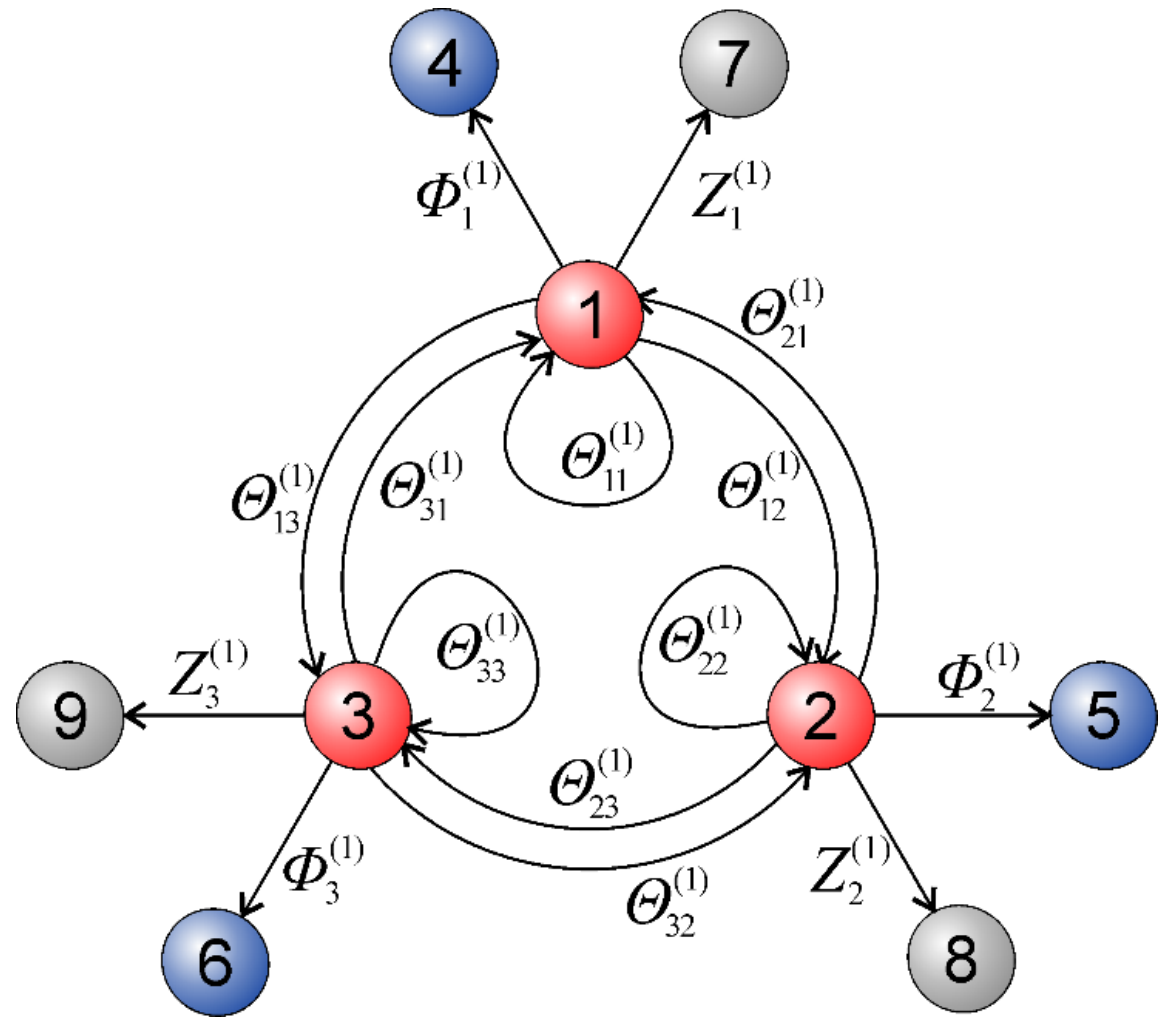

**Figure 3.** Nine-state Markov process diagram showing states, elementary events, and their SPDFs in a three-component system. The red circles illustrate the states of transient localization of MEE on molecules of component 1, 2, and 3, respectively. The blue circles and gray circles illustrate the terminating states in which MEE is instantaneously converted to either photons or heat, respectively. The symbols $\Theta^{(1)}_{ij}$ , $\Phi^{(1)}_i$ , and $Z^{(1)}_i$ denote the SPDFs describing the MEE transitions indicated by the arrows.

We compute the desired elements of the matrix $\hat{\Phi}(s)$ using expression (90). In this way, we obtain

$$\left.\begin{aligned} \hat{\Phi}_{ii}(s) &= \frac{1}{D}\left[\left(1-\hat{\Theta}^{(1)}_{jj}\right)\left(1-\hat{\Theta}^{(1)}_{kk}\right) - \hat{\Theta}^{(1)}_{jk}\hat{\Theta}^{(1)}_{kj}\right]\hat{\Phi}^{(1)}_i \\ \hat{\Phi}_{ij}(s) &= \frac{1}{D}\left[\hat{\Theta}^{(1)}_{ij}\left(1-\hat{\Theta}^{(1)}_{kk}\right) + \hat{\Theta}^{(1)}_{ik}\hat{\Theta}^{(1)}_{kj}\right]\hat{\Phi}^{(1)}_j \end{aligned}\right\} \quad i, j, k = \{1,2,3\}, \quad k \neq j \neq i \tag{167}$$

Expressions describing elements of the matrix $\phi$ are found using relation (110), which gives

$$\left.\begin{aligned} \phi_{ii} &= \frac{1}{d}\left[\left(1-\theta^{(1)}_{jj}\right)\left(1-\theta^{(1)}_{kk}\right) - \theta^{(1)}_{jk}\theta^{(1)}_{kj}\right]\phi^{(1)}_i \\ \phi_{ij} &= \frac{1}{d}\left[\theta^{(1)}_{ij}\left(1-\theta^{(1)}_{kk}\right) + \theta^{(1)}_{ik}\theta^{(1)}_{kj}\right]\phi^{(1)}_j \end{aligned}\right\} \quad i, j, k = \{1,2,3\}, \quad k \neq j \neq i \tag{168}$$

where

$$
\begin{aligned}
d = &\left(1-\theta_{11}^{(1)}\right)\left(1-\theta_{22}^{(1)}\right)\left(1-\theta_{33}^{(1)}\right) \\
&-\left(1-\theta_{11}^{(1)}\right)\theta_{23}^{(1)}\theta_{32}^{(1)}-\left(1-\theta_{22}^{(1)}\right)\theta_{13}^{(1)}\theta_{31}^{(1)} \\
&-\left(1-\theta_{33}^{(1)}\right)\theta_{12}^{(1)}\theta_{21}^{(1)}-\theta_{12}^{(1)}\theta_{23}^{(1)}\theta_{31}^{(1)}-\theta_{13}^{(1)}\theta_{32}^{(1)}\theta_{21}^{(1)}
\end{aligned}
\tag{169}
$$

The values of the parameters $\theta_{ij}^{(1)}$ and $\phi_i^{(1)}$ can be found using the equations (83) and (84), respectively. When NET processes are not reversible in the system under study, the expressions for $\hat{\Phi}(s)$ and $\phi$ simplify significantly. For example, considering the condition $\hat{\Theta}_{ij}^{(1)}(s)=0$ when $j<i$ causes the matrix (168) to reduce to the form

$$
\phi=\begin{bmatrix} K_1 & K_{12}K_2 & (K_{13}+K_{12}K_{23})K_3 \\ 0 & K_2 & K_{23}K_3 \\ 0 & 0 & K_3 \end{bmatrix} \tag{170}
$$

where

$$
K_i=\frac{\phi_i^{(1)}}{1-\theta_{ii}^{(1)}} \tag{171}
$$

$$
K_{ij}=\frac{\theta_{ij}^{(1)}}{1-\theta_{ii}^{(1)}} \tag{172}
$$

The last three expressions are in agreement with the analogous expressions given in the paper [24]

### 4.3 *Example of the application of TRMC and TUMC methods to the analysis of absorbing and partially absorbing states*

So far, we have considered the application of the TRMC method to study the properties of chains that do not contain absorbing states. However, it turns out that this method can also provide interesting results in the case of chains that do contain such states. For example, consider a two-state Markov chain in which state 1 is both transient and partially absorbing, and state 2 is absorbing. We will call state *i* an absorbing state if $P_{ii}^{(1)}=1$, and we will call state *i* a partially absorbing state if $0<P_{ii}^{(1)}<1$. In both cases, $P_{ii}^{(1)}$ is the probability of transition of the chain object from state *i* to state *i*, related to the time-dependent function $\mathbb{P}_{ii}^{(1)}(t)$ through expression (54). A diagram of this chain is shown in Fig. 4.

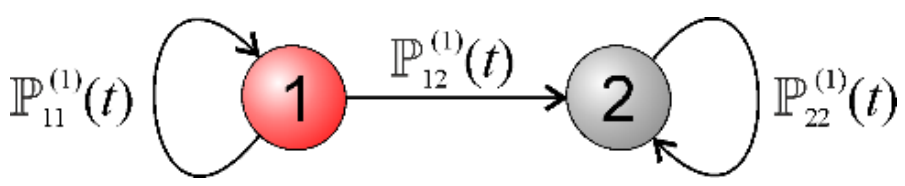

**Figure 4.** Two-state Markov process diagram adopted for the time analysis of the absorbing state. State 1 is both a transient and partially absorbing state and state 2 is an absorbing state. The functions $\mathbb{P}_{11}^{(1)}(t)$, $\mathbb{P}_{12}^{(1)}(t)$, and $\mathbb{P}_{22}^{(1)}(t)$ are SPDFs of the transitions from state 1 to state 1, from state 1 to state 2, and from state 2 to state 2, respectively

To simplify our discussion, we will assume that the one-step temporal probability density distributions of the transitions in this system are exponential, characterized by positive rate constants $k_{11}$, $k_{12}$, and $k_{22}$. The SPDF matrix of one-step transitions used in TRMC description therefore has the form

$$
\mathbb{P}^{(1)}(t)=\begin{bmatrix} k_{11}e^{-(k_{11}+k_{12})t} & k_{12}e^{-(k_{11}+k_{12})t} \\ 0 & k_{22}e^{-k_{22}t} \end{bmatrix} \tag{173}
$$

Hence, the matrix $\hat{\mathbb{P}}^{(1)}(s)$ is given by

$$
\hat{\mathbb{P}}^{(1)}(s)=\begin{bmatrix} \dfrac{k_{11}}{s+k_{11}+k_{12}} & \dfrac{k_{12}}{s+k_{11}+k_{12}} \\ 0 & \dfrac{k_{22}}{s+k_{22}} \end{bmatrix} \tag{174}
$$

After inserting (174) into Eq. (74), we obtain a version of matrix $\mathbb{P}$ dependent on *s*

$$
\hat{\mathbb{P}}(s)=\begin{bmatrix} \dfrac{s+k_{11}+k_{12}}{s+k_{12}} & \dfrac{k_{12}(s+k_{22})}{s(s+k_{12})} \\ 0 & \dfrac{s+k_{22}}{s} \end{bmatrix} \tag{175}
$$

which, when inverted to time space, takes the form

$$
\mathbb{P}(t)=\begin{bmatrix} \delta(t)+k_{11}e^{-k_{12}t} & k_{12}e^{-k_{12}t}+k_{22}(1-e^{-k_{12}t}) \\ 0 & \delta(t)+k_{22} \end{bmatrix} \tag{176}
$$

The $\mathbb{P}_{22}(t)$ element of the $\mathbb{P}(t)$ matrix represented by $\delta(t)+k_{22}$ determines the temporal course of the visit intensity to state 2 after it has been excited at time $t=0$. As can be seen, this intensity does not decrease over time, but remains constant. This can be understood such a way that, due to the absorbing properties of state 2, a type of continuously repeating excitation (CRE) with a repetition frequency $k_{22}$ arises in this state. The form of element $\mathbb{P}_{12}(t)$ of matrix (176) shows that the same type of CRE is generated in state 2 after excitation from state 1. This is indicated by the fact that for long times, the value of element $\mathbb{P}_{12}(t)$ remains constant and equal to $k_{22}$. The repetition rate of excitation in state 2 can be reduced by decreasing the value of the parameter $k_{22}$. This results in a decrease in the intensity of CRE until it is completely extinguished when $k_{22}=0$. However, setting $k_{22}=0$ in the absence of energy transfer from state 2 to other states means that state 2 is transformed into a terminating state. We can see that the terminating state can also be understood as a limiting version of the absorbing state, when the repetition rate in the latter state tends to zero.

The one-step residence probability matrix calculated based on the equations (15) and (17) takes the form

$$
\mathcal{P}^{(1)}(t)=\begin{bmatrix} e^{-(k_{11}+k_{12})t} & 0 \\ 0 & e^{-k_{22}t} \end{bmatrix} \tag{177}
$$

After inserting matrices (175) and $\mathcal{L}$-transform of (177) into Eq. (117), we find the residence probability matrix

$$
\mathcal{P}(t)=\begin{bmatrix} e^{-k_{12}t} & 1-e^{-k_{12}t} \\ 0 & 1 \end{bmatrix} \tag{178}
$$

We can see here that despite the presence of $k_{11}$ and $k_{22}$ in equations (175) and (177), expression (178) does not depend on $k_{11}$ and/or $k_{22}$. Hence, we conclude that expression (178) also holds when $k_{11}=0$ and/or $k_{22}=0$. It is easy to verify that $\mathcal{P}(t)$ given by this expression satisfies Chapman-Kolmogorov equation (123). This follows from the fact that all elements of matrix (173) are expressed by memoryless functions.

Within the TUMC/DTMC framework, the one-step transition probability matrix $P^{(1)}$ is of fundamental importance. In our conditions, it can be easily calculated either by integrating the (173) matrix over time from zero to infinity, or by setting $s = 0$ in the (174) matrix. This way, we obtain

$$P^{(1)} = \begin{bmatrix} \dfrac{k_{11}}{k_{11} + k_{12}} & \dfrac{k_{12}}{k_{11} + k_{12}} \\ 0 & 1 \end{bmatrix} \tag{179}$$

An attempt to calculate the elements of the transition probability matrix $P$ using expressions (99) and (179) fails because the $(I_v - P^{(1)})$ matrix becomes singular. However, the matrix $P$ can also be found from matrix (175) as $\hat{\mathbb{P}}(s = 0)$ or by integrating matrix (176) over time from zero to infinity. This gives

$$P = \begin{bmatrix} \dfrac{k_{11} + k_{12}}{k_{12}} & \infty \\ 0 & \infty \end{bmatrix} \tag{180}$$

The infinite values of the number of transfers to state 2 result from the assumed absorbing properties of this state. We can also see that increasing the value of parameter $k_{11}$ or decreasing the value of parameter $k_{12}$ increases the number of transfers to state 1 when state 1 is initially excited. This is related to the fact that state 1 is thus given increasingly pronounced absorbing properties. To wrap up this section, let's point out that the result shown in expression (180) could not be found by using the TUMC/DTMC method alone. To find it, we integrated the result of the previously applied TRMC method over time.

## 5 Conclusions

The main achievements of this work can be listed as follows

1. To determine the effect of RNET on the values of parameters characterizing the fluorescence of MCS containing $n$ components, we used a suitably adapted Markov chain technique. To the best of our knowledge, the Markov chain technique has not yet been used for this purpose.
2. In our considerations, the DTMC model was subject to adaptation. We concluded that for the description of the MEE transfer, it is advisable to introduce a new type of Markov process object state, which we called the terminating state. We subsequently demonstrated the necessity of this modification in Section 4.3. A characteristic feature of the terminating state is that, once it is reached, the Markov process object loses the ability to move to any of the states in the chain. The object of our Markov process is MEE. The introduction of terminating states enables the inclusion of events such as the emission of a fluorescence photon or an internal conversion in a given process. It is also important for the computational process that the replacement of absorbing states by terminating states makes matrices of the type $I_n - P^{(1)}$ invertible.
3. To find the time characteristics of the fluorescence emission probabilities of the individual MCS components, we introduced a new type of Markov chain, which we called the time-resolved Markov chain (TRMC). What is significant for TRMC is that the individual rows of its transition matrix are filled not by univariate discrete probability distributions, but by bivariate probability mass-density functions. Unlike CTMC, in our TRMC model the time distributions of events possible in each state need not be exponential, and can be arbitrary.
4. Within TRMC, the step number $\mu$ is independent of real time. The time distributions of the time random variable $T$ considered at different values of $\mu$ are mutually independent. This fact is important in qualifying whether a given process is Markovian or not. We consider any TRMC that satisfies the modified Markov property (52) to be Markovian, regardless of whether the time distributions of events at each step of the process are exponential or not.
5. In our opinion, contrary to its name, the commonly used DTMC method does not include time. We argue that in DTMC, time is unspecified, hence we propose renaming DTMC as TUMC.
6. Under TRMC, time is always continuous. On the other hand, discrete or continuous can be the temporal probability density distributions of the chain object's transition between states. A variant of TRMC using continuous time is obtained if the time distributions of the probability densities of one-step transitions between states of the chain object are continuous. A variant using discrete time can be obtained by assuming that the time distributions of these probability densities are discrete, e.g., such as that defined by Eq. (126).
7. Unlike CTMC, using TRMC, transitions within the same state can be taken into account. In the case of MCS, this makes it possible, for example, to determine the effect of MEE migration on the fluorescence intensity of individual components.
8. In Section 3.9, we have shown that Equation (74), which is the basic equation of the TRMC method, is consistent with the analogous equation derived earlier in the RT. However, our TRMC approach differs from the RT approach in that it refers to the SPDFS of individual events, not to their SCDFs, as is the case in RT. This allows the widely described Laplace transform technique to be used in the TRMC method instead of the rarely described Laplace-Stieltjes transform technique used by RT.
9. In Section 3.10, we have shown that the results obtained using the TRMC method are equivalent to those obtained using conventional analytical methods. However, the TRMC method is simpler and more perceptual. We have also shown that the previously known CTMC method is a special case of our TRMC method. The two methods become equivalent when the temporal probability density distributions of transitions between process states are described by exponential (memoryless) functions.
10. In this work, the TRMC method has been successfully applied to describe the course of a relatively narrow range of phenomena involving MCS fluorescence. In our opinion, the range of applications of this method can be much broader. Undoubtedly, there is a need for further work on both its applications and its theoretical basis.

## Acknowledgements

The author is pleased to acknowledge helpful suggestions and comments from Prof. Joachim Domsta.

## Appendix A: Exponentialization of fluorescence decays of individual components in the absence of heterotransfer

In the case where there is only homotransfer in a given MCS and no heterotransfer, the $\Theta^{(1)}(t)$ matrix becomes a diagonal matrix. It follows from Eq. (87) that the matrix $\hat{\Theta}(s)$ is then also diagonal, and its non-zero elements $\hat{\Theta}_{ii}(s)$ are of the form

$$\hat{\Theta}_{ii}(s) = \left(1 - \hat{\Theta}^{(1)}_{ii}(s)\right)^{-1} \tag{181}$$

From Eq. (90), we can see that, under these conditions, the matrix $\hat{\Phi}(s)$ containing the Laplace transforms of the fluorescence decay functions of the individual MCS components is also diagonal. It is easy to see that the non-zero elements of this matrix are given by the expression

$$\hat{\Phi}_{ii}(s) = \frac{\hat{\Phi}^{(1)}_{i}(s)}{1 - \hat{\Theta}^{(1)}_{ii}(s)} \tag{182}$$

Assuming that $\Theta^{(1)}_{ii}(t)$ and $\Phi^{(1)}_{i}(t)$ are described by Eqs. (23) and (26), respectively, and that in Eq. (17) $\mathbb{K}_{ij}(t) = 0$ for $j \neq i$, we can write

$$\hat{\Phi}_{ii}(s) = \frac{\left(\phi_{0i}/\tau_{0i}\right)\int_0^\infty g_i(t)\, h_i(t)\, dt}{1 + \int_0^\infty g_i(t)\, dh_i(t)} \tag{183}$$

where

$$g_i(t) = \exp\left[-\left(s + \tau_{0i}^{-1}\right)t\right] \tag{184}$$

$$h_i(t) = \exp\left[-\int_0^t \mathbb{K}_{ii}(t')\, dt'\right] \tag{185}$$

Calculating the integral in the denominator of expression (183) by parts, we find that

$$\hat{\Phi}_{ii}(s) = \frac{\phi_{0i}}{\tau_{0i}} \frac{1}{s + \tau_{0i}^{-1}} \tag{186}$$

It follows that in the absence of heterotransfer, the time course of the original function from transform (182) is exponential and described by the expression

$$\Phi_{ii}(t) = \frac{\phi_{0i}}{\tau_{0i}} \exp\left(-\frac{t}{\tau_{0i}}\right) \tag{187}$$

This is despite the fact that the time courses of the functions $\Theta^{(1)}_{ii}(t)$ and $\Phi^{(1)}_{i}(t)$ are not exponential. For example, this occurs under conditions where Eqs. (42) and (43) reduce to the form of

$$\Theta^{(1)}_{ii}(t) = \frac{\gamma_{ii}}{\sqrt{\tau_{0i}\, t}}\, g(t;\gamma_{ii},\tau_{0i}) \tag{188}$$

$$\Phi^{(1)}_{i}(t) = \frac{\phi_{0i}}{\tau_{0i}}\, g(t;\gamma_{ii},\tau_{0i}) \tag{189}$$

where the values of the function $g(t;\gamma_{ii},\tau_{0i})$ can be calculated using Eq. (41).